\documentclass[usegraphicx]{mn2e}

\usepackage{mathptmx}
\usepackage{amssymb}
\usepackage{amsbsy}
\usepackage{url}
\usepackage{helvet}

\newcommand{\Mpc}{\rm\thinspace Mpc}
\newcommand{\kpc}{\rm\thinspace kpc}

\newcommand{\km}{\rm\thinspace km}

\newcommand{\cm}{\rm\thinspace cm}

\newcommand{\pcmcu}{\hbox{$\cm^{-3}\,$}}

\newcommand{\s}{\rm\thinspace s}

\newcommand{\keV}{\rm\thinspace keV}

\newcommand{\kmps}{\hbox{$\km\s^{-1}\,$}}

\newcommand{\kmpspMpc}{\hbox{$\kmps\Mpc^{-1}$}}

\newcommand{\Zsun}{\hbox{$\thinspace \mathrm{Z}_{\odot}$}}

\begin{document}
\title[Contour binning]{Contour binning: a new technique for
  spatially-resolved X-ray spectroscopy applied to Cassiopeia A}

\author[J.S. Sanders]{J.S. Sanders\thanks{E-mail: jss@ast.cam.ac.uk}\\
  Institute of Astronomy, Madingley Road, Cambridge. CB3 0HA}
\maketitle

\begin{abstract}
  We present a new technique for choosing spatial regions for X-ray
  spectroscopy, called ``contour binning''. The method chooses regions
  by following contours on a smoothed image of the object. In addition
  we re-explore a simple method for adaptively smoothing X-ray images
  according to the local count rate, we term ``accumulative
  smoothing'', which is a generalisation of the method used by
  \textsc{fadapt}. The algorithms are tested by applying them to a
  simulated cluster data set.  We illustrate the techniques by using
  them on a 50~ks \emph{Chandra} observation of the Cassiopeia A
  supernova remnant. Generated maps of the object showing abundances
  in eight different elements, absorbing column density, temperature,
  ionisation timescale and velocity are presented.  Tests show that
  contour binning reproduces surface brightness considerably better
  than other methods. It is particularly suited to objects with
  detailed spatial structure such as supernova remnants and the cores
  of galaxy clusters, producing aesthetically pleasing results.
\end{abstract}


\section{Introduction}
Many X-ray telescopes such as \emph{XMM-Newton}, \emph{Chandra} and
\emph{ROSAT} contain detectors which capture individual photons,
recording their energy and position on the detector. Therefore unlike
conventional optical observing techniques, we simultaneously collect
imaging and spectroscopic data for each part of the object. In
addition the \emph{Chandra X-ray Observatory} has very high spatial
resolution ($\sim 1$~arcsec), allowing the properties of the emitting
object to be studied in unprecedented detail.

With this spectroscopic and imaging information we can select
``events'' from the observation corresponding to a particular part of
an object with a ``region filter''. From these events a spectrum can
be built-up. An X-ray spectral package such as \textsc{xspec} (Arnaud
1996) can then be used to fit a physical model to the spectrum.
Conventionally simple geometric shapes such as annuli, sectors, boxes
or ellipses are used to define a region filter. Using sectors, for
example, and assuming spherical or elliptical symmetry, one can
account for projection in a cluster of galaxies. However most extended
objects are not symmetric when observed in detail (for example, the
Perseus cluster [Fabian et al 2000; Sanders et al 2004], the
Cassiopeia-A supernova remnant [Hwang et al 2004], and Abell 2052
[Blanton, Sarazin \& McNamara 2003]).

Given the morphological diversity of extended X-ray sources it is
important to have techniques which allow us to analyse the spectral
variation of a source over its extent. We first investigated this
problem when looking for cool gas in a sample of X-ray clusters using
\emph{ROSAT} PSPC archival data (Sanders, Fabian and Allen 2000). We
devised a technique which used adaptively smoothed maps (Ebeling et
al, in preparation) to define contours in surface brightness. The
ratio of the number of counts in different energy bands between each
contour was used to define an X-ray colour. By using a grid of models,
the absorption and temperature of the gas could be estimated between
the contours.

We approached this problem again with the advent of data from
\emph{Chandra} with its high spatial resolution. We created an
algorithm called ``adaptive binning'' (Sanders \& Fabian 2001) which
used the uncertainty on the number of counts or the error on the ratio
of counts in different bands to define the size of binning region
used. The process was simple: pass over the image, copying those
pixels which have a small enough uncertainty on the number of counts
or colour to an output image. Bin up the remaining pixels by a factor
of two. Repeat until all the pixels have been binned. On the final
pass we bin any pixels which are not yet binned. This simple approach
works well and was used by us and other authors on data from several
clusters of galaxies (e.g.  Centaurus -- Sanders \& Fabian 2002;
Perseus -- Fabian et al 2000; Abell~4059 --- Choi et al 2004).

The disadvantage of this approach is that the binning scale varies by
factors of two. It is very noticeable where the scale changes, and
some regions are overbinned. Therefore we started using the ``bin
accretion'' algorithm of Cappellari \& Copin (2003). The algorithm
adds pixels to a bin until a signal-to-noise threshold is reached.
After all the pixels have been accreted, it uses Voronoi tessellation
to make tessellated regions based on the weighted position of the
original bins. This technique has the advantage of creating bins which
are compact, varying in size smoothly with the surface brightness, and
also provides bins with similar signal-to-noise ratios.  We applied the
method to X-ray observations of the Perseus cluster (Sanders et al
2004) and Abell~2199 (Johnstone et al 2002).  Rather than use X-ray
colours, we extracted spectra for each of the regions and used
spectral fitting to derive, for example, temperature and abundance
maps. Recently Diehl \& Statler (2006) have generalised this algorithm
to allow for data whose signal-to-noise does not add in quadrature.

The motivation for further work in this area is that the methods above
do not use the surface brightness distribution to change the shape of
each bin.  Physical parameters (e.g. density, temperature and
abundance) usually change in the direction of surface brightness
changes.  The method we describe here uses the surface brightness to
define bins which cover regions of similar brightness.

Other methods have been presented for mapping the parameters of the
intracluster medium. These included wavelet techniques (Bourdin et al
2004) and monte carlo methods (Peterson, Jernigan \& Kahn 2004). The
advantage of binning techniques is that they provide errors on
individual spectral fit parameters, or colours, from a particular part
of the sky. The individual measurements made using binning methods are
independent, making it easy to easily measure the significance of
individual spatial features.

The techniques presented in this paper have already been applied to a
number of \emph{Chandra} observations of clusters, including a deep
observation of the complex structure of the Centaurus cluster (Fabian
et al 2005), the possible detection of nonthermal radiation and the
identification of a high metal shell likely to be associated with a
fossil radio bubble in the Perseus cluster (Sanders et al 2005), a
sample of moderate redshift clusters (Bauer et al 2005), and a 900~ks
observation of the Perseus cluster (Fabian et al 2006), finding little
evidence for temperature changes associated with shock-like features,
and producing evidence of a substantial reservoir of cool X-ray
emitting material.

We first present a simple smoothing method (``accumulative
smoothing''), and then present the binning method based on the
smoothed image (``contour binning'').

\section{Accumulative smoothing}
In order to bin using the surface brightness it was necessary for us
to get an estimate of the surface brightness in an image in the
absence of noise and counting statistics.  Simple Gaussian smoothing
can be sufficient, but if the brightness of the object varies over its
extent the smoothing scale will be too small or too large in parts of
the image.  There are several methods for adaptively smoothing an
image based on the surface brightness (e.g. \textsc{asmooth}, Ebeling
et al, in preparation; \textsc{csmooth} in \textsc{ciao}, based on
\textsc{asmooth}; Huang \& Sarazin 1996). We present a method called
accumulative smoothing, which is a generalisation of the method used
by the \textsc{ftools} routine \textsc{fadapt}, now including handling
background and exposure variation.  In addition we adapt the method to
create smoothed colour maps. This smoothing method has the advantage
of being fast, simple and easy to interpret.  It allows the use of
blank-sky background images (rather than trying to estimate the local
background), masks, and exposure maps. Its disadvantage is that it
does not guarantee to globally preserve counts. This is probably not a
serious problem as it is usually more important to preserve the local
count rate. We will discuss how well it does locally later.

It is a very simple routine which smooths an image with a top-hat
kernel, whose size varies as a function of position. The size is
varied so that the kernel contains a minimum signal-to-noise ratio.

Suppose there is an input image $\textbf{\textsf{I}}$, containing
numbers of counts in each pixel There is also a background image
$\textbf{\textsf{B}}$ (optionally fixed to zero).  The smoothed count
image at positions $\mathbf{r}$, where $\mathbf{r}$ is a vector
representing the coordinates of the centre of the pixel being
examined, is
\begin{equation}
  \textbf{\textsf{S}}(\mathbf{r}) = \frac{1}{N}
  \sum_{| \mathbf{a}-\mathbf{r} | \le R(\mathbf{r})}
  { \left[ \textbf{\textsf{I}}(\mathbf{a}) - \frac{E_i}{E_b} \textbf{\textsf{B}}(\mathbf{a})
    \right] },
\end{equation}
where $N$ is the number of pixels summed over ($\sim \pi
R[\mathbf{r}]^2$), and the smoothing radius at the pixel,
$R(\mathbf{r})$, is defined in Equation~\ref{eqn:radius}. $E_b$ and
$E_i$ are the exposure times of the foreground and background
observations, respectively. If we wish to take account of the
variation in exposure times over an image, we can create a smoothed
count rate image,
\begin{equation}
  \textbf{\textsf{S}}_\mathrm{rate}(\mathbf{r}) = \frac{1}{N}
  \sum_{| \mathbf{a}-\mathbf{r} | \le R(\mathbf{r})}
      { \left[ \frac{\textbf{\textsf{I}}(\mathbf{a})}{E_i(\mathbf{a})}
          -
          \frac{\textbf{\textsf{B}}(\mathbf{a})}{E_b(\mathbf{a})}
      \right] },
\end{equation}
The smoothing radius $R(\mathbf{r})$ is defined to be the minimum
value of the radius (in integer units) where the signal-to-noise is
greater or equal to a threshold value. Summing over pixels at
positions $\mathbf{a}$, where $| \mathbf{a}-\mathbf{r} | \le
R(\mathbf{r})$, then the condition is met when
\begin{equation}
  [S/N](\mathbf{r}) \sim
  \frac{ 
    \sum
    \left[
      \textbf{\textsf{I}}(\mathbf{a}) -
      \frac{E_i(\mathbf{a})}{E_{b}(\mathbf{a})} \textbf{\textsf{B}}(\mathbf{a})
    \right]
    }{
    \left(
      g \left[
        \sum
        \textbf{\textsf{I}}(\mathbf{a})
      \right]
      + 
      \frac{1}{N}
      \sum
      \left\{ \left[
        \frac{E_i(\mathbf{a})}{E_b(\mathbf{a})}
      \right]^2 \right\}
      g \left[
        \sum
        \textbf{\textsf{B}}(\mathbf{a})
      \right]
    \right)^{1/2}
  }
  \ge M.
  \label{eqn:radius}
\end{equation}
$R(\mathbf{r})$ is found by incrementing $R(\mathbf{r})$ from 1 until
it is true. $g(c)$ is an approximation for the squared-uncertainty on
$c$ counts, obtained from equation 7 of Gehrels (1986) when $S=1$,
\begin{equation}
  g(c) = \left( 1 + \sqrt{c + \frac{3}{4}} \right)^2.
\end{equation}
This approximation is the larger error bar on $c$ counts, so we
overestimate the negative error by assuming the errors are symmetric.

Equation~\ref{eqn:radius} is approximate if the ratio of the exposure
time of the image and background varies over the image. We do not sum
the squared error of $\textbf{\textsf{B}}(a)$ as the sum of the
approximation for the uncertainty squared $g(c)$ will become
increasingly inaccurate with more terms.

Therefore the algorithm smooths the image using a top-hat kernel with
a variable smoothing radius. The smoothing radius varies so that it
contains a minimum signal-to-noise ratio. We handle edges and masked
regions by ignoring pixels outside the valid region ($N$ does not
include these pixels).

\subsection{Tests of the algorithm}
\label{sect:acsmtest}
\begin{figure*}
  \includegraphics[width=0.3\textwidth]{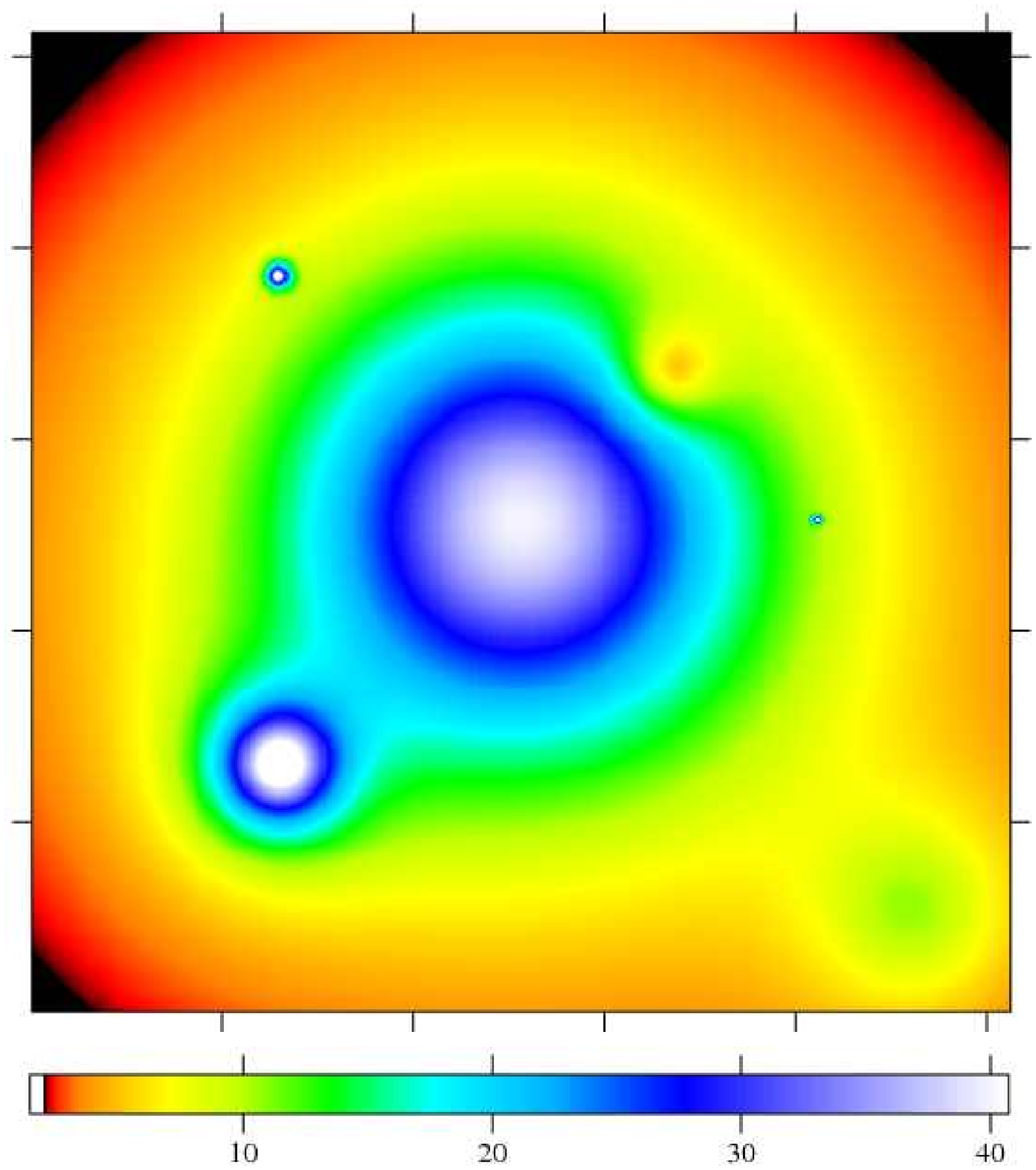}
  \includegraphics[width=0.3\textwidth]{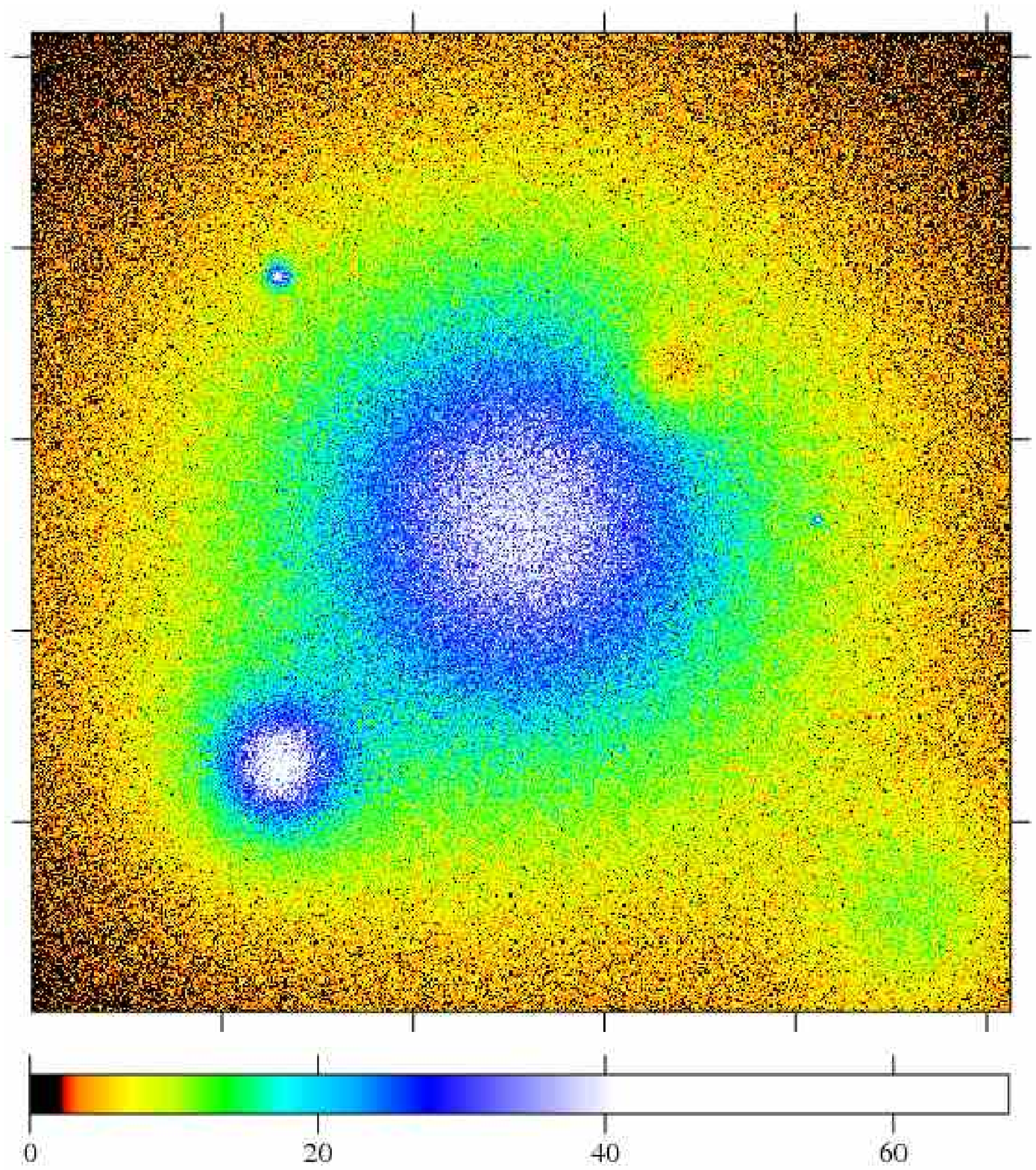}
  \includegraphics[width=0.3\textwidth]{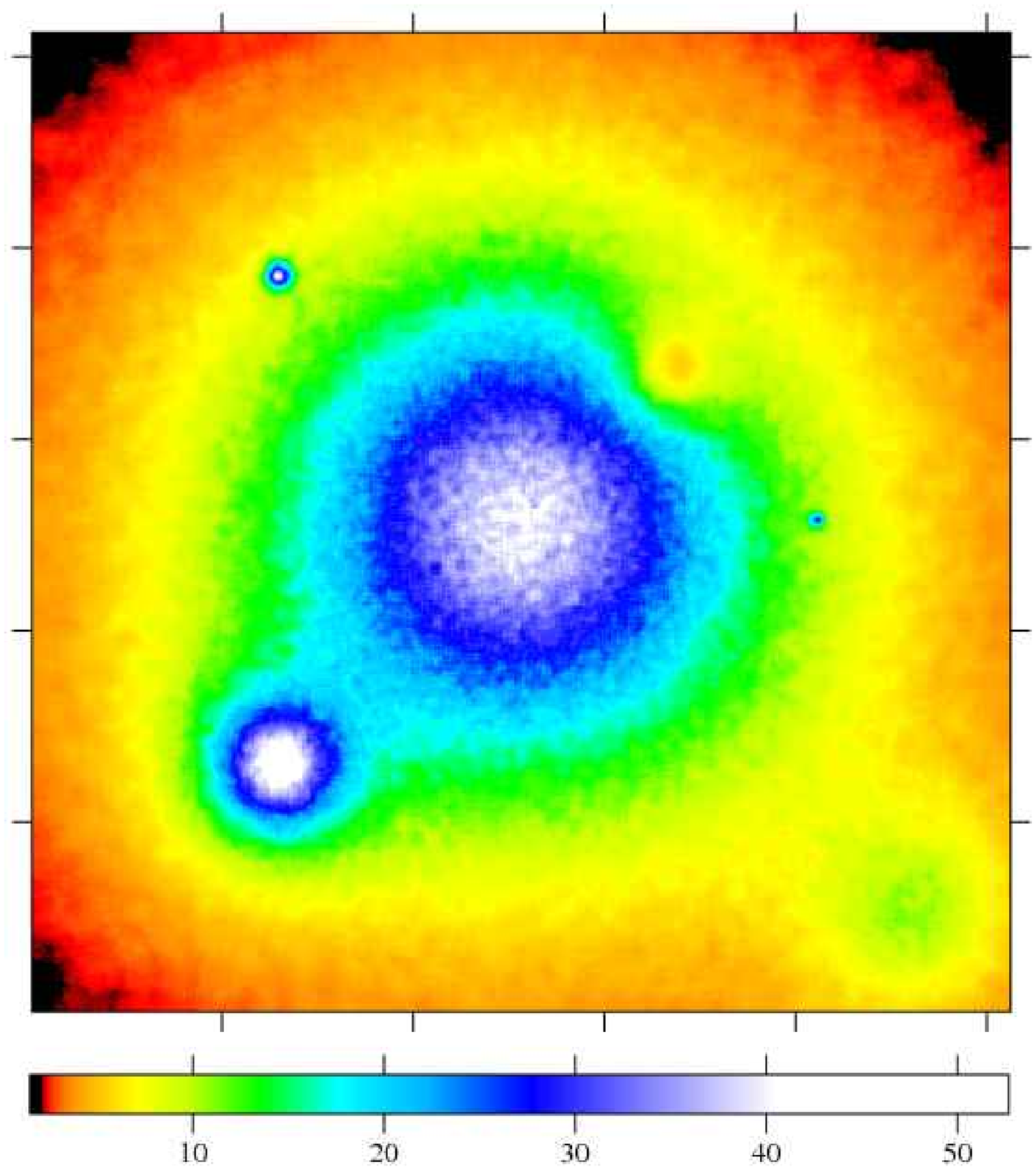}
  \caption{(Left) Surface brightness model made up of six $\beta$
    model components. (Middle) Realisation of the model assuming
    Poisson statistics. (Right) Accumulatively smoothed image of the
    realisation, with a signal-to-noise ratio of 20. All three images
    use the same colour scale. The units are counts per pixel.}
  \label{fig:sb}
\end{figure*}

To test how closely the smoothing algorithm reproduces the surface
brightness of a model image, we constructed a surface brightness model
made up of six $\beta$ surface brightness model components, where each
component had surface brightness
\begin{equation}
S = S_0 \left(1 + \left[\frac{r}{r_c}\right]^2 \right) ^{-3\beta+
  1/2}.
\label{eqn:beta}
\end{equation}
The components are listed in Table~\ref{tab:beta}. We constructed a
model $512 \times 512$ pixel image based on the components
(Fig.~\ref{fig:sb}~[left]). A Poisson statistic realisation of the
model is shown in Fig.~\ref{fig:sb}~(centre). In
Fig.~\ref{fig:sb}~(right) we include an accumulatively smoothed image
of the Poisson realisation.

\begin{table}
  \begin{tabular}{lllll}
    $S_0$ (counts) & $\beta$ & $r_c$ (fraction) & $x_c$ (fraction) &
    $y_c$ (fraction)  \\ \hline
    40 & 0.67 & 1/4 & 1/2 & 1/2 \\
    40 & 0.67 & 1/16 & 1/4 & 1/4 \\
    40 & 1 & 1/64 & 1/4 & 3/4 \\
    40 & 5 & 1/64 & 0.8 & 0.5 \\
    8 & 0.67 & 1/16 & 0.9 & 0.1 \\
    -12 & 0.67 & 1/16 & 0.65 & 0.65 \\
    
  \end{tabular}
  \caption{Surface brightness components. Each is a $\beta$ surface
    brightness model (Equation \ref{eqn:beta}). The core radius
    ($r_c$), component $x$ centre ($x_c$) and $y$ centre ($y_c$) are expressed
    as a fraction of the $512\times512$ pixel image. The coordinates
    are measured from the lower left of the images.}
  \label{tab:beta}
\end{table}

\begin{figure*}
  \includegraphics[width=0.247\textwidth]{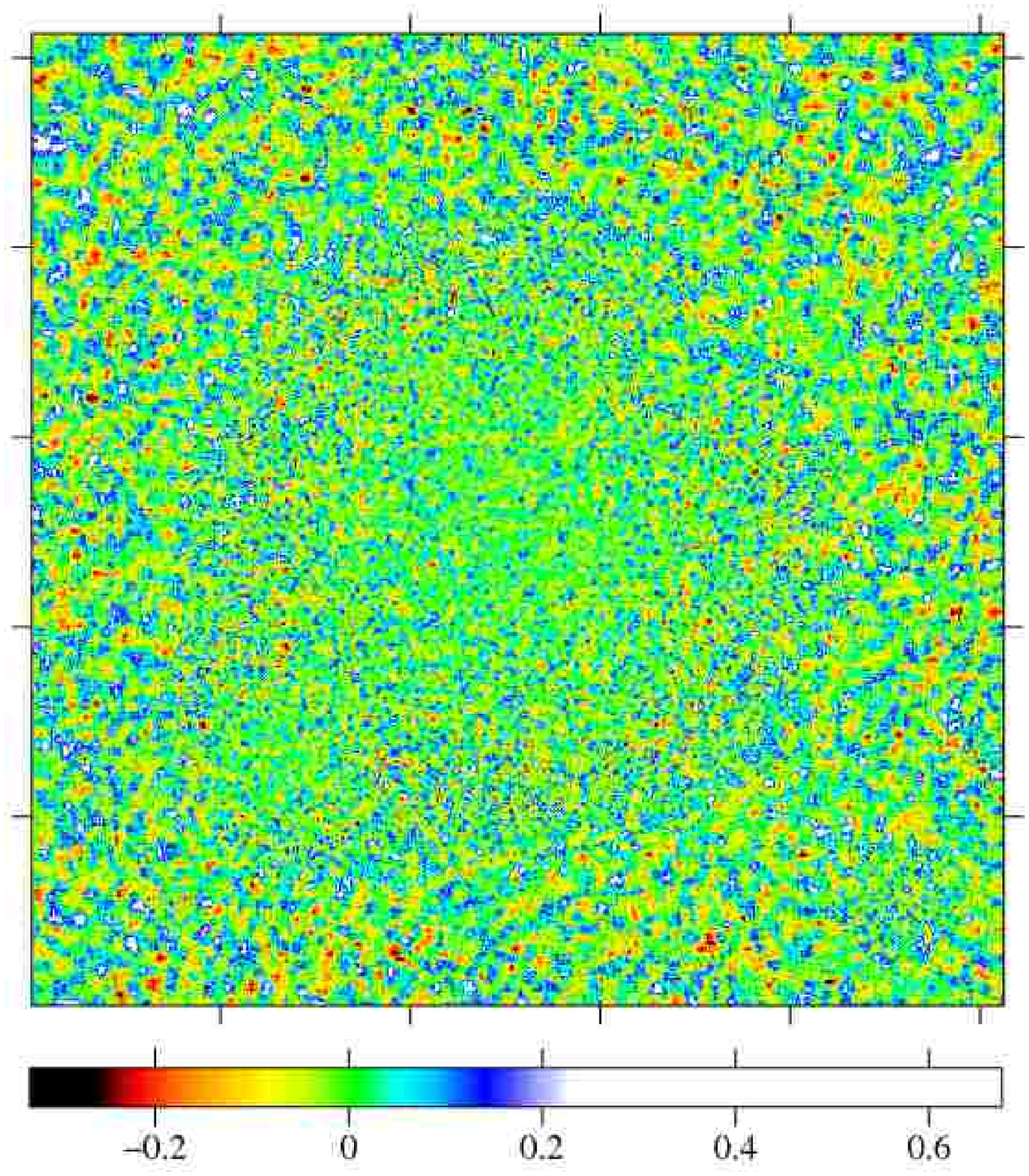}
  \includegraphics[width=0.247\textwidth]{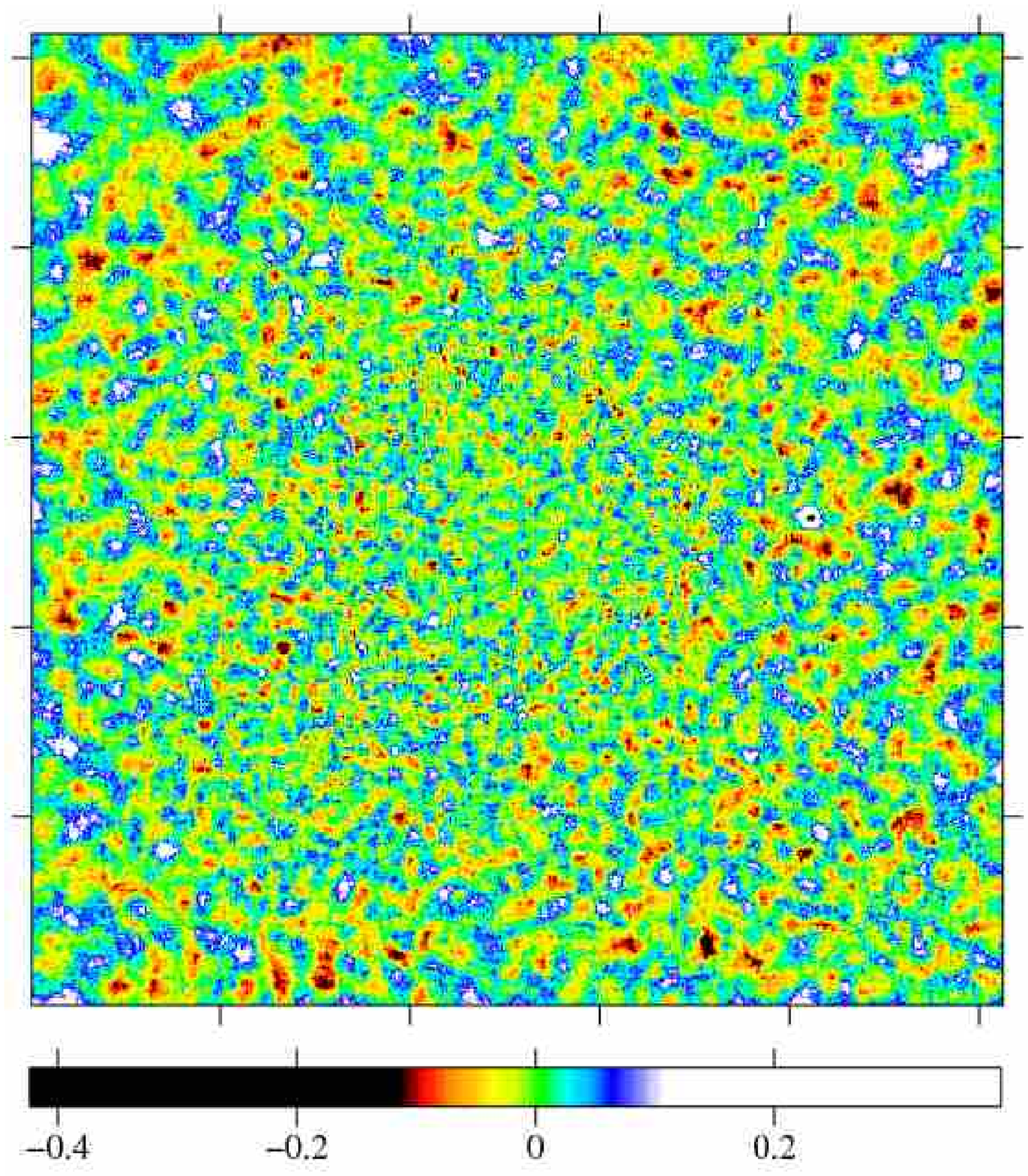}
  \includegraphics[width=0.247\textwidth]{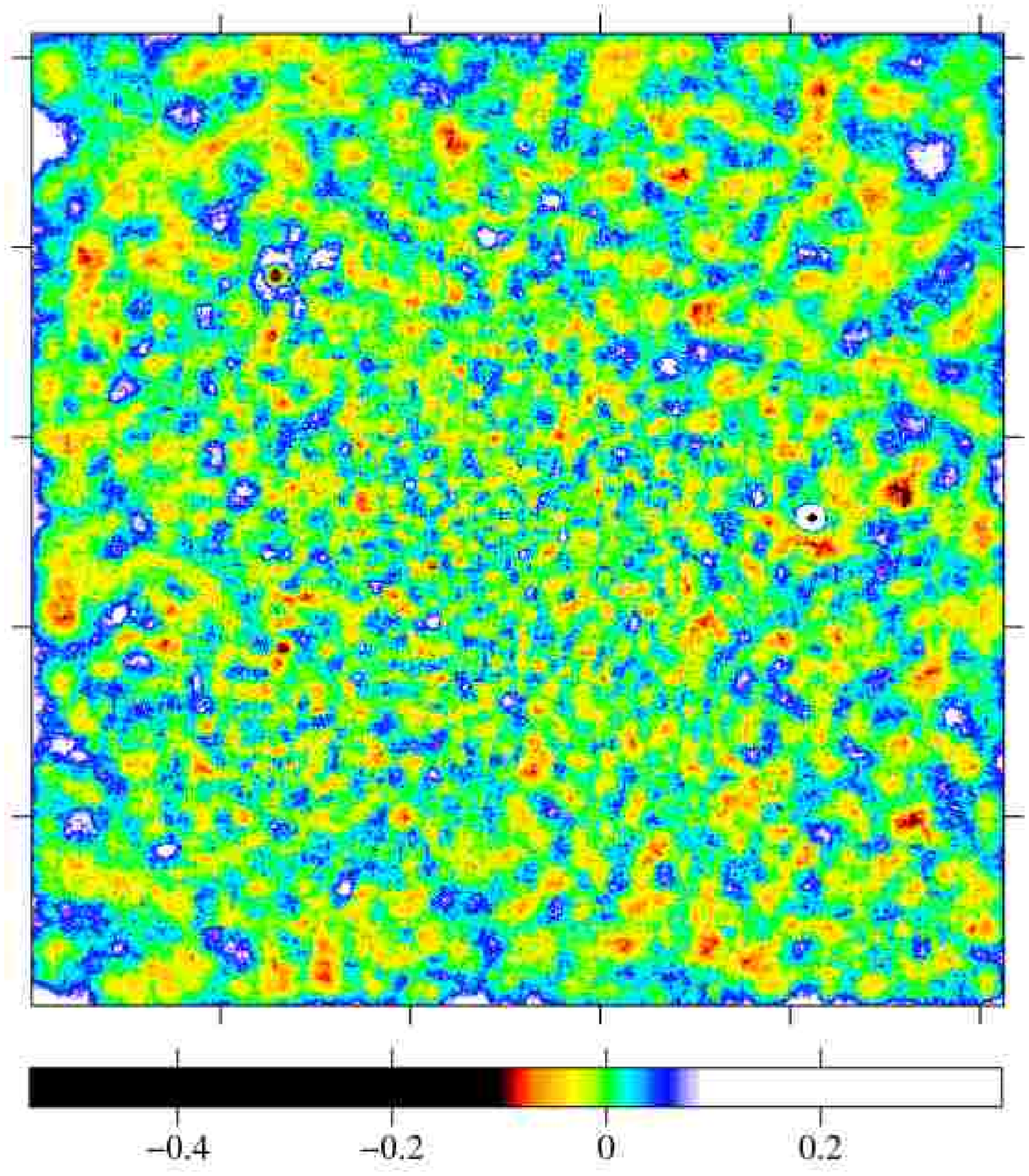}
  \includegraphics[width=0.247\textwidth]{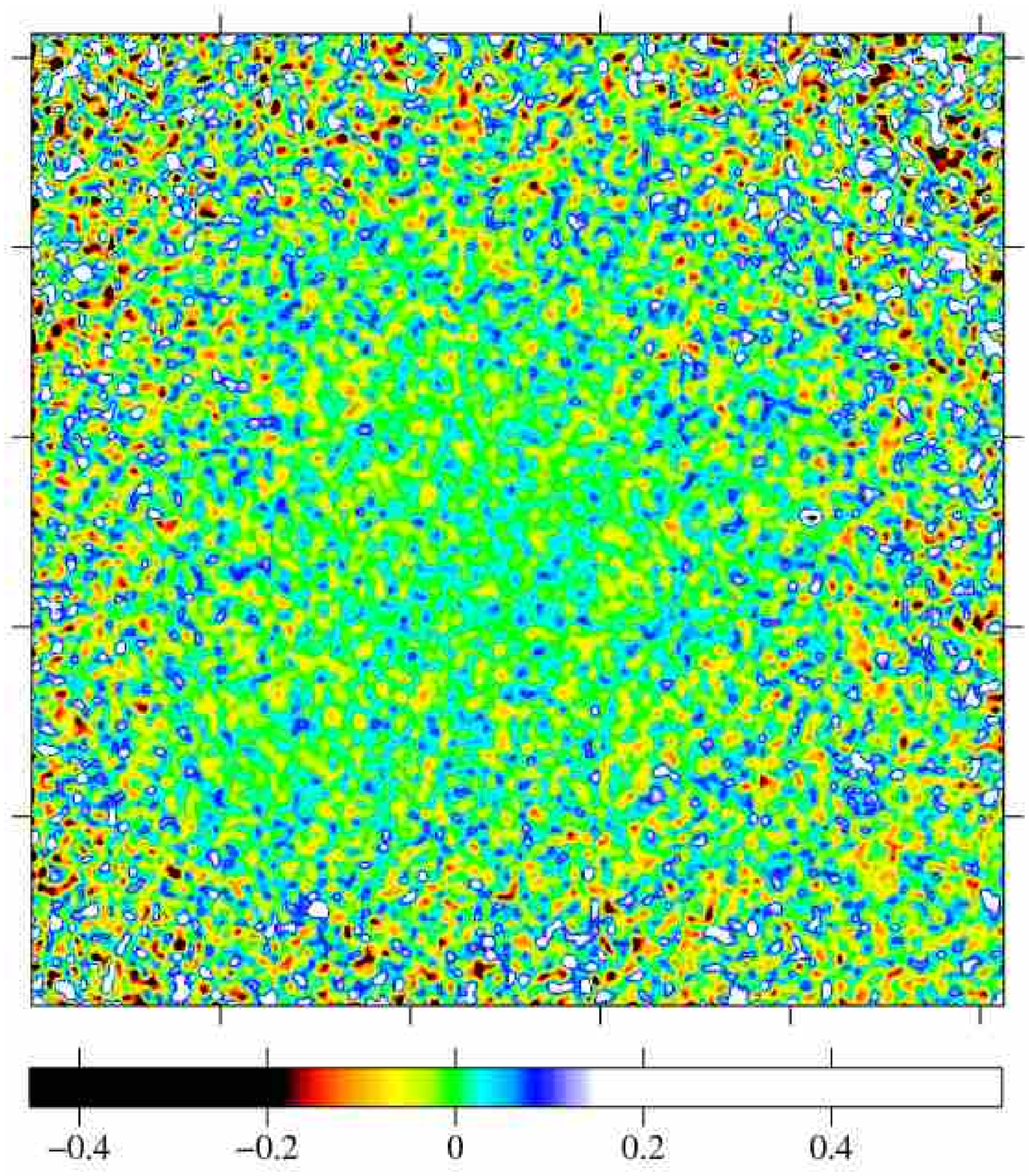}
  \caption{Fractional difference between complex cluster model and
    accumulatively-smoothed image, for $S/N$ 8, 20 and 30, and using
    Gaussian smoothing with a smoothing scale of 2 pixels. Note that
    each image uses a different colour scale to show the magnitude of
    the scatter.}
  \label{fig:sb_deltas}
\end{figure*}

\begin{figure}
  \includegraphics[width=\columnwidth]{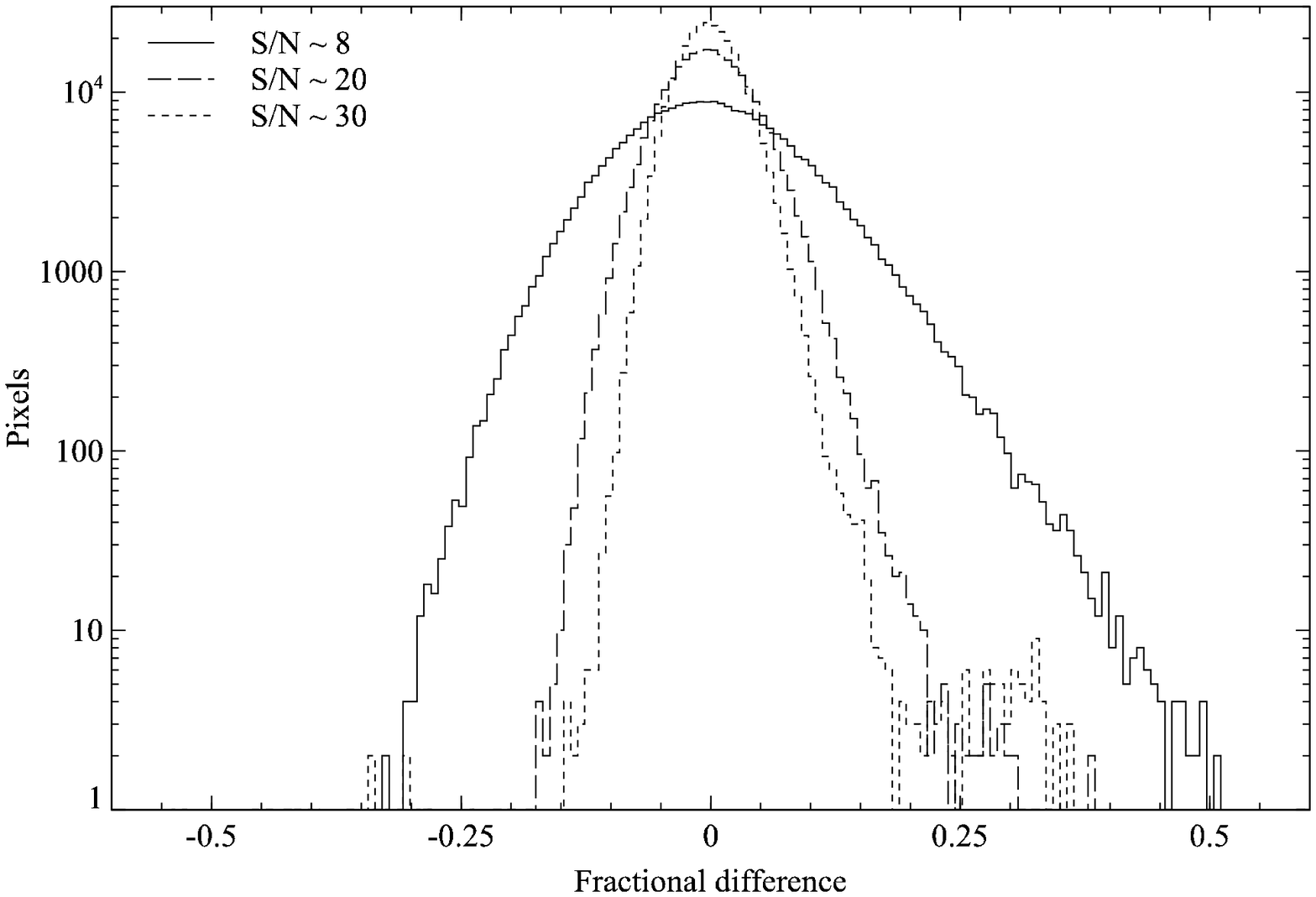} \vspace{2mm} \\
  \includegraphics[width=\columnwidth]{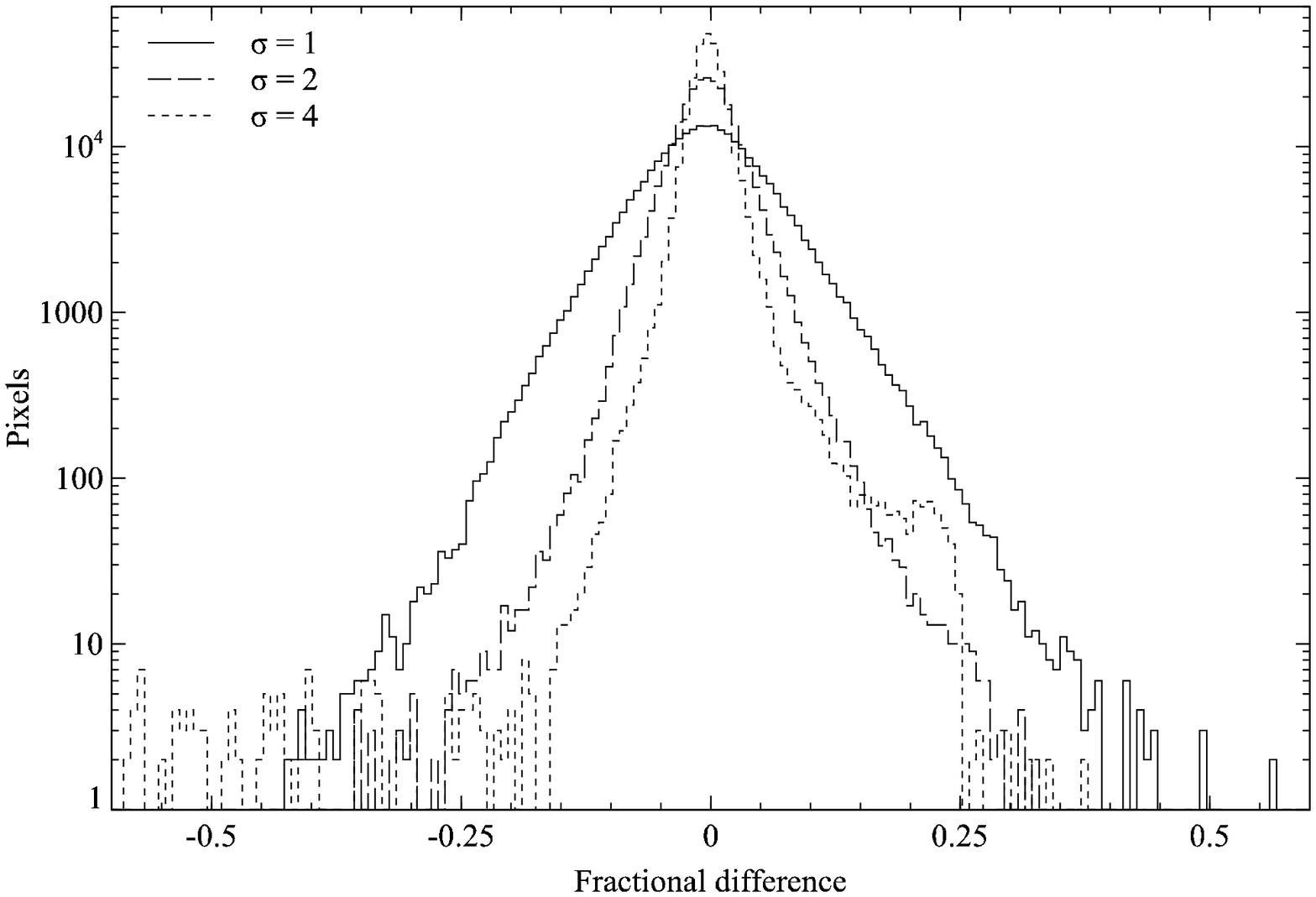} \vspace{2mm} \\
  \includegraphics[width=\columnwidth]{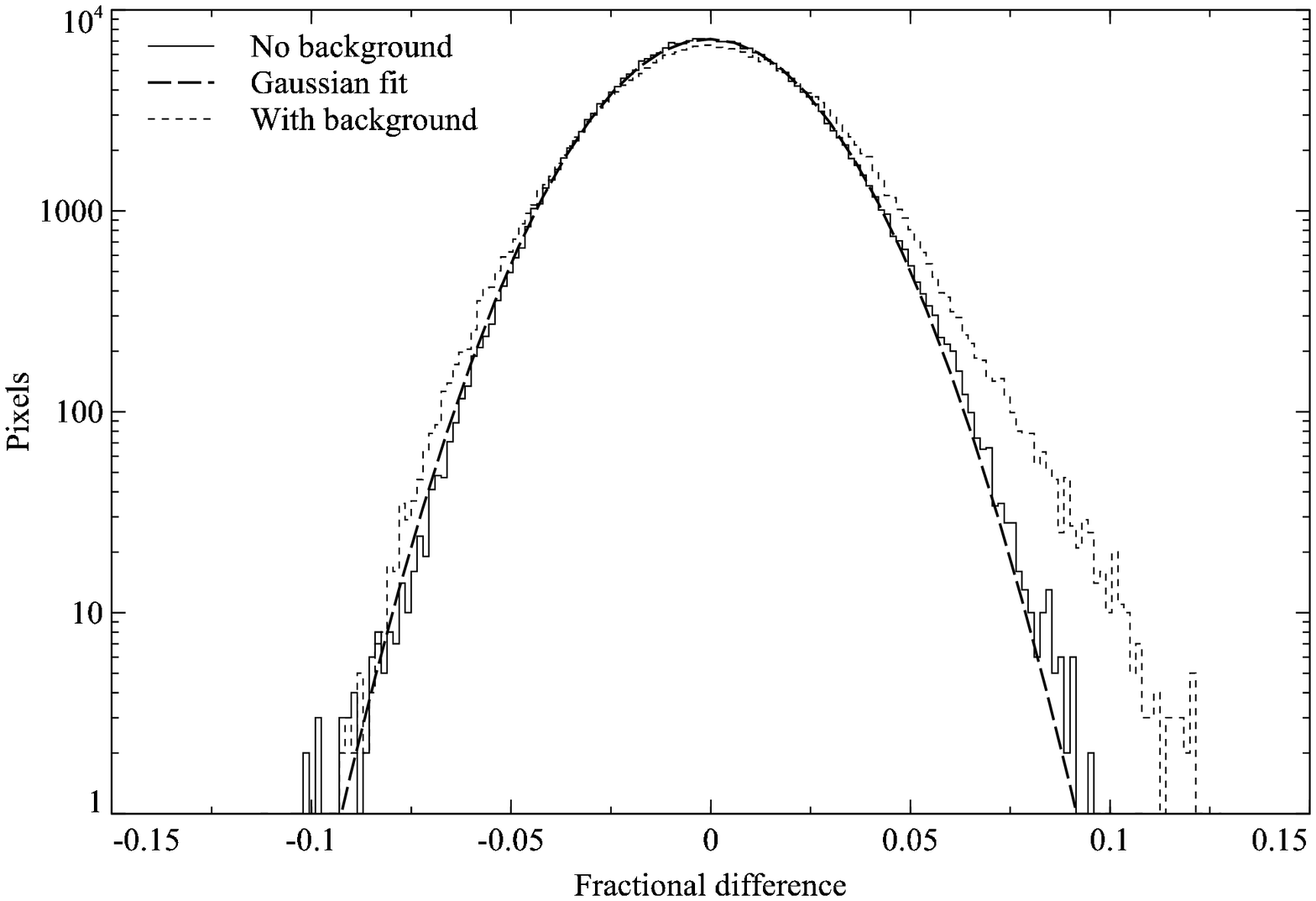}
  \caption{Histogram of fractional differences between reconstructed
    surface brightness and original model. (Top) Difference between
    complex cluster model and accumulatively-smoothed data. The three
    curves are smoothed with signal-to-noise threshold ratios of 30,
    20 and 8.  (Centre) Fractional difference between the model and
    Gaussian-smoothed data (using widths of 1, 2 and 4 pixels).
    (Bottom) Differences between a single-component $\beta$ model and
    accumulative smoothed data (signal-to-noise threshold of 40). The
    smooth line is a Gaussian fit to the differences, with a best
    fitting width of 0.023. The edges of the image have been removed
    to eliminate edge effects. Also plotted is the result when a
    constant background is added to the model.}
  \label{fig:deltahist}
\end{figure}

A good smoothing algorithm should reproduce the model image with no
large scale deviations or biases, and random small scale deviations.
We computed the fractional difference between smoothed images of the
realisation of the model with various signal-to-noise threshold ratios
(Fig.~\ref{fig:sb_deltas}). Also shown is a Gaussian smoothed
fractional difference map for comparison.  To aid comparison we show
histograms of the fractional differences using accumulative smoothing
and Gaussian smoothing with different signal-to-noise ratios and scale
lengths in Fig.~\ref{fig:deltahist}.

Firstly, accumulative smoothing mostly shows deviations from the model
which do not vary in magnitude over the image. The Gaussian smoothed
deviations in Fig.~\ref{fig:sb_deltas}~(right) increase when the count
rate decreases. The central part of the image is uniform in colour,
while the outer parts alternate between the extremes. Where
accumulative smoothing shows deviations are at the edge of the image
and around centrally-concentrated sources when the smoothing scale is
too large (e.g.  Fig.~\ref{fig:sb_deltas}~[centre right]). Near the
edges the surface brightness is too high, as there are no lower flux
pixels further out to smooth over. If a centrally-concentrated source
does not have enough signal to exceed the signal-to-noise threshold,
$M$, then it is smoothed into its surroundings. The flux at the centre
of the source is underestimated, and the flux in its surroundings is
overestimated.

We took a single $\beta$ component surface brightness image (using an
image of size $1024\times 1024$ pixels, and applied Gaussian
($\sigma=8$ pixels using the \textsc{fgauss} program from
\textsc{ftools}) and accumulative smoothing ($M=40$) to a Poisson
realisation of the model. In the centre of the image, the Gaussian
smoothing length is approximately the accumulative smoothing length.

The deviation distributions (Fig.~\ref{fig:deltahist}) show that
accumulative smoothing produces narrower tails in the distribution
than Gaussian smoothing. The distributions can be well fit with a
Gaussian, except for a tail in the positive deviations. At low
signal-to-noise thresholds this may be due to the approximation in the
uncertainties of the counts as symmetric. The Gehrels approximation
used overestimates the uncertainty on low numbers of counts, meaning
that faint regions will have larger kernels applied than necessary
when a low signal-to-noise threshold is used.  At high signal-to-noise
the tail is due to edge effects and oversmoothed compact sources.
Using a model distribution that contains just the central $\beta$
component, and excluding the edges of the image, produces a
distribution which is almost Gaussian (Fig.~\ref{fig:deltahist}
[bottom]). The width of the distribution is 0.023, which is very close
to a simple estimation of the width ($1 / M = 0.025$). To get the
estimate, we require $c = M^2$ counts in the absence of a background
to exceed the threshold. The standard deviation on a mean of $c$
counts is $\sqrt{c}$, or as a fraction of the mean, $1/\sqrt{c} =
1/M$.

We have also considered the case of a simple $\beta$ model with a
background. We added a background of 40 counts across the entire image
(which is the same as the peak of the $\beta$ distribution). The
distribution of deviations (Fig.~\ref{fig:deltahist} [bottom]) is
fairly Gaussian, although there is a significant tail at larger
deviations than the case without a background. In this test we assumed
the background was very well determined (simulated by using a much
larger exposure time for the background than the foreground).

\subsection{Colour maps}
Colour maps show the ratio of counts in two bands. Such maps are
useful because trends in X-ray colour can follow physical trends (e.g.
temperature or absorption). Accumulative smoothing can also be used to
generate smoothed colour maps. Smoothing or binning is very useful in
creating colour maps, as the error on a ratio of two counts can
greatly vary over an X-ray image.

Rather than considering a threshold on the signal-to-noise in the
smoothing kernel, a maximum uncertainty on the X-ray colour can be
used instead.

\subsection{Conclusions}
It can be concluded that accumulative smoothing, and the smoothing of
\textsc{fadapt} when there is no background or exposure variation,
does not introduce a major bias to an image. The median difference is
negligible for all three smoothing thresholds shown. Accumulative
smoothing is certainly better than Gaussian smoothing at smoothing on
appropriate scales over an image.  Indeed the deviation range does not
change with position. It also produces a distribution of deviations
that shows a narrow, almost Gaussian, tail, although the tail is wider
for positive deviations at low signal-to-noise. If there is not enough
signal in point sources they are smoothed into their surroundings,
leading to deviations from the intrinsic surface brightness of the
sky. In comparison Gaussian smoothing shows much more noise in regions
where the count rate is low compared to where it is high.

In addition the algorithm is particularly suited for presentation of
X-ray images as an alternative to Gaussian smoothing. It may well be
better in many cases to other forms of adaptive smoothing as it
introduces few artifacts. 

\section{Contour binning}
Here we describe how the accumulatively smoothed map can be used to
define independent spatial regions. These regions can be used for
spectral extraction and fitting, to measure physical parameters, or
for the calculation of X-ray colours. The method need not apply to
accumulatively smoothed images. Other techniques such as adaptive
smoothing could be used instead.

Simply, the algorithm, starting at the highest flux pixel on a
smoothed image, adds pixels nearest in surface brightness to a bin
until the signal-to-noise ratio exceeds a threshold algorithm. A new
bin is then created. This technique naturally creates bins which
follow the surface brightness.

\subsection{Initial pass}
Contour binning takes the smoothed image,
$\textbf{\textsf{S}}(\mathbf{r})$. We start from the highest flux
pixel in the smoothed image.  The pixel is added to the set of pixels
in the current bin, $\beta$. If the signal-to-noise of the pixels in
$\beta$ is greater or equal than a binning threshold $T$,
\begin{equation}
  [S/N]_{\beta} \sim
  \frac{
  \sum_{ \mathbf{a} \in \beta }
      { \left[ \textbf{\textsf{I}}(\mathbf{a}) -
          \frac{E_i(\mathbf{a})}{E_{b}(\mathbf{a})} \textbf{\textsf{B}}(\mathbf{a})
      \right] }
  } {
    \left( \sum_{ \mathbf{a} \in \beta }
    { \left\{ g[\textbf{\textsf{I}}(\mathbf{a})] +
        \left[\frac{E_i(\mathbf{a})}{E_{b}(\mathbf{a})} \right]^2
        g[\textbf{\textsf{B}}(\mathbf{a})] \right\} } \right)^{1/2}
  }
  \ge T,
\end{equation}
then the region is complete. Otherwise we iterate over the subset of
$\beta$ which lie at the edge of the bin. We consider all the unbinned
neighbours of these pixels and take the one which has a flux in the
smoothed image closest to the starting flux of the bin.  This pixel is
added to $\beta$, and we repeat the process. If there are no
neighbouring unbinned pixels we stop this bin. We consider a pixel to
be a neighbour if it differs in one of its coordinates by 1, and in
the others by 0 (diagonally neighbouring pixels could also be
included, however).

Once we have completed the bin, we start binning again. The starting
pixel is again the highest flux pixel in the smoothed image which has
not yet been binned. To optimise this potentially expensive search, we
sort the pixels in the smoothed image into flux order before binning. 

\subsection{Cleaning pass}
After all the pixels have been binned, there will be some bins with a
signal-to-noise ratio of less than $T$. Many of these will contain
single pixels which were left stranded between other bins. The number
of these stranded bins decreases with increased smoothing of the
initial image.  We therefore ``clean up'' the bins to remove any below
the threshold. We start by selecting the bin with the lowest
signal-to-noise ratio.  By examining its edge pixels, we select the
pixel which is closest in smoothed value to a pixel in a neighbouring
bin.  This pixel is moved into the corresponding neighbouring bin
(thereby increasing the signal-to-noise ratio of its neighbour). We
repeat, transferring the pixels until there are no pixels remaining in
the lowest signal-to-noise bin. We move onto the next remaining bin
which has the lowest signal-to-noise and repeat the whole process. The
cleaning pass ends when there are no more bins with a signal-to-noise
ratio of less than the threshold $T$.

\subsection{Constraints}
\label{sect:constrain}
On a high signal-to-noise dataset with low signal-to-noise bins the
above process is sufficient. However there are possible undesirable
features under other conditions. If the smoothed map is not very
smooth, the bins can become distended (this can be improved at the
detriment of potential detail by increasing $M$). If the smoothed
map is very smooth and radial, then bins will become whole annuli if
$T$ is too large.  In some cases this outcome is desirable.

To prevent these problems we can impose extra geometric constraints
during the binning process. Ideally such a constraint could be
computed in constant time on each iteration or else growing large bins
becomes prohibitively expensive. We identified a fast constraint that
gave the most aesthetic results. The procedure is to estimate a radius
$R$ that the current bin would have if it were a circle, i.e. $R \sim
\sqrt{ N / \pi}$, where $N$ is the number of pixels in the bin. By
examining circles on a pixelated grid, we calculate this value exactly
for integer radii. This is far better than the simple estimate for
small circles. We then measure the distance $d$ of the pixel to be
added from the current flux-weighted centroid of the bin.  If $R/d >
C$, where $C$ is the constraint parameter, then we do not add the
pixel.  The constraint is also applied during the cleaning pass by
moving pixels into only bins in which the constraint would be
satisfied. If there are no bins which would satisfy the constraint, the
constraint is broken to prevent isolated pixels.

\subsection{Tests of the algorithm}
\begin{figure*}
  \includegraphics[width=0.247\textwidth]{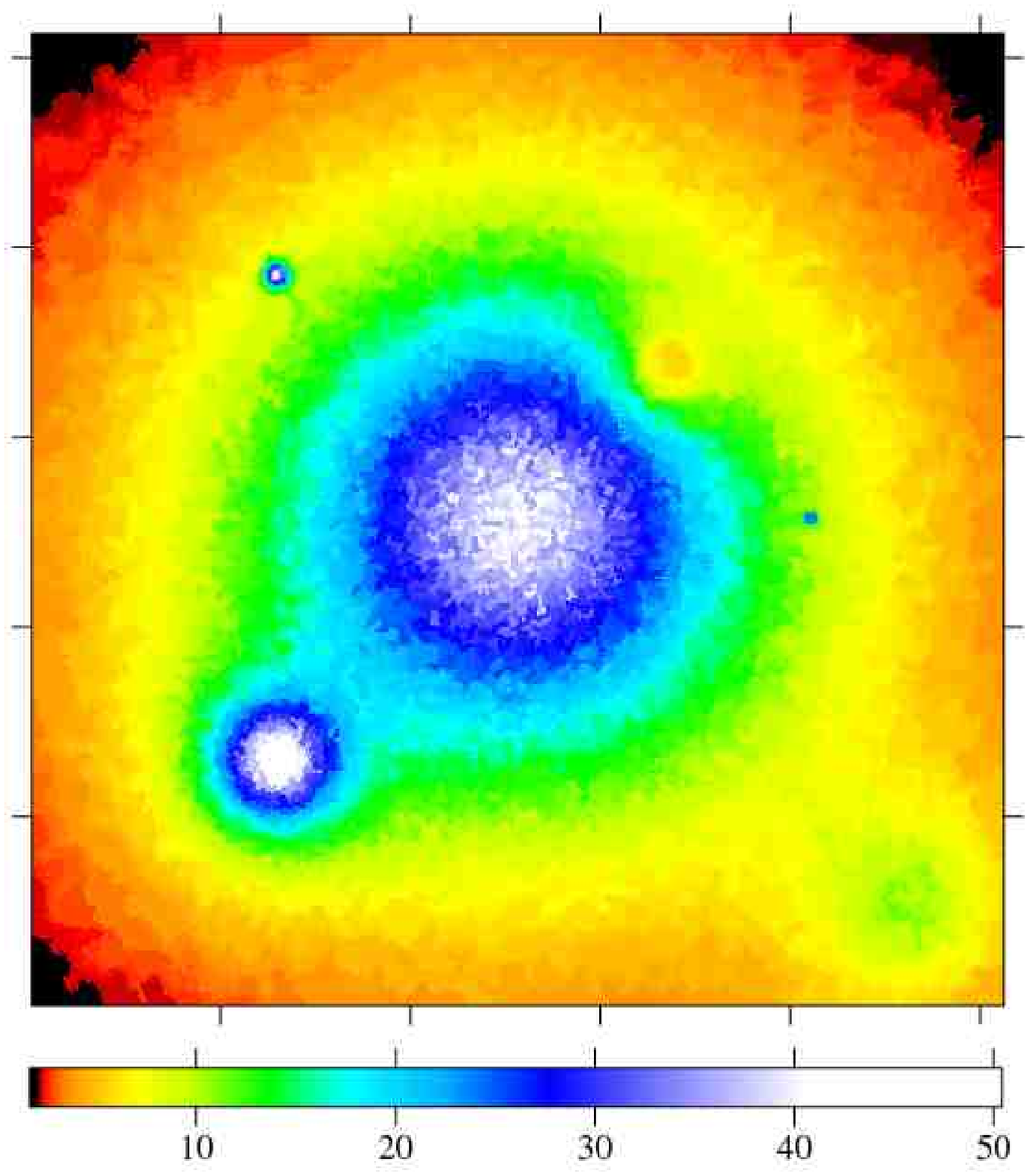}
  \includegraphics[width=0.247\textwidth]{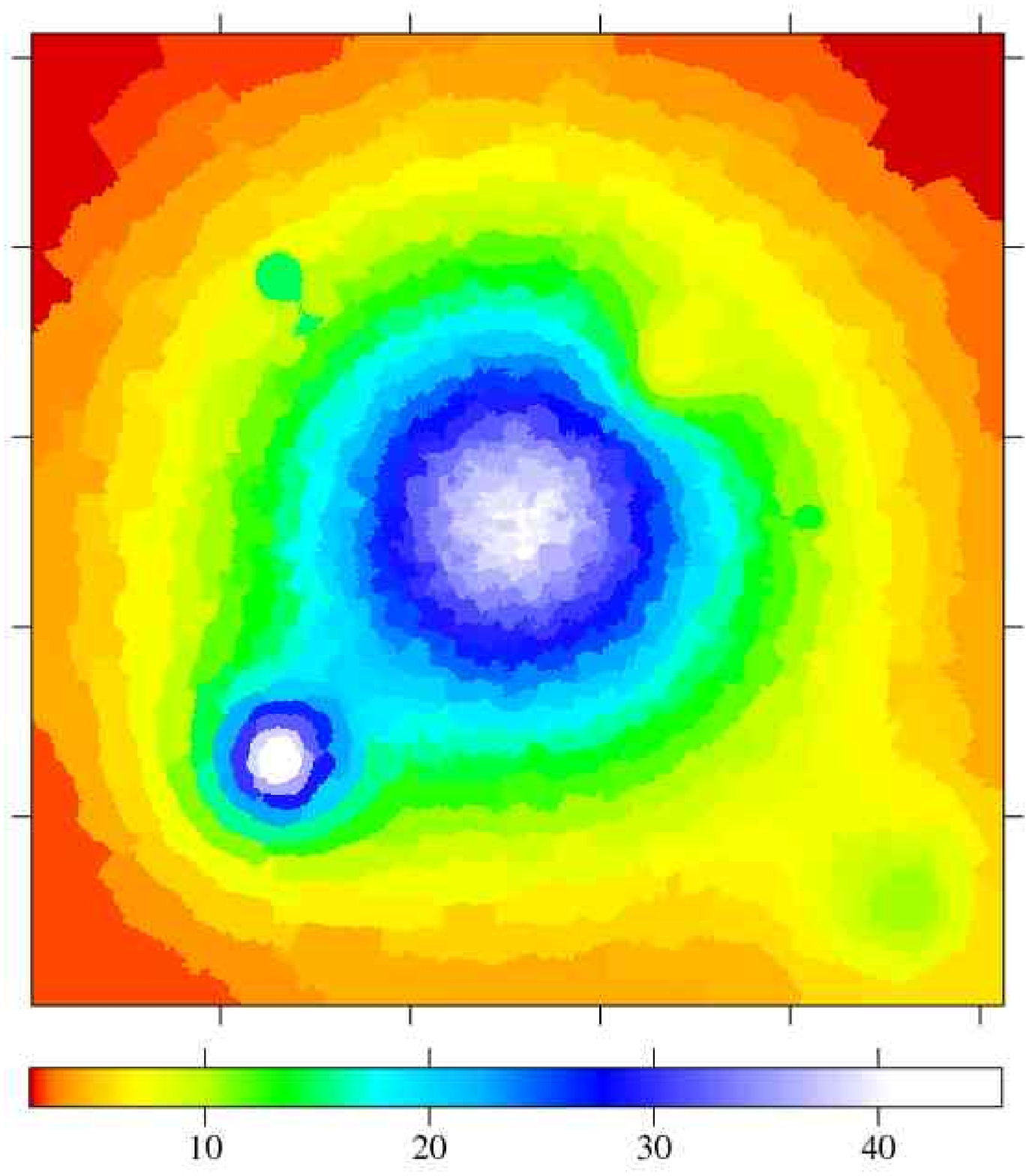}
  \includegraphics[width=0.247\textwidth]{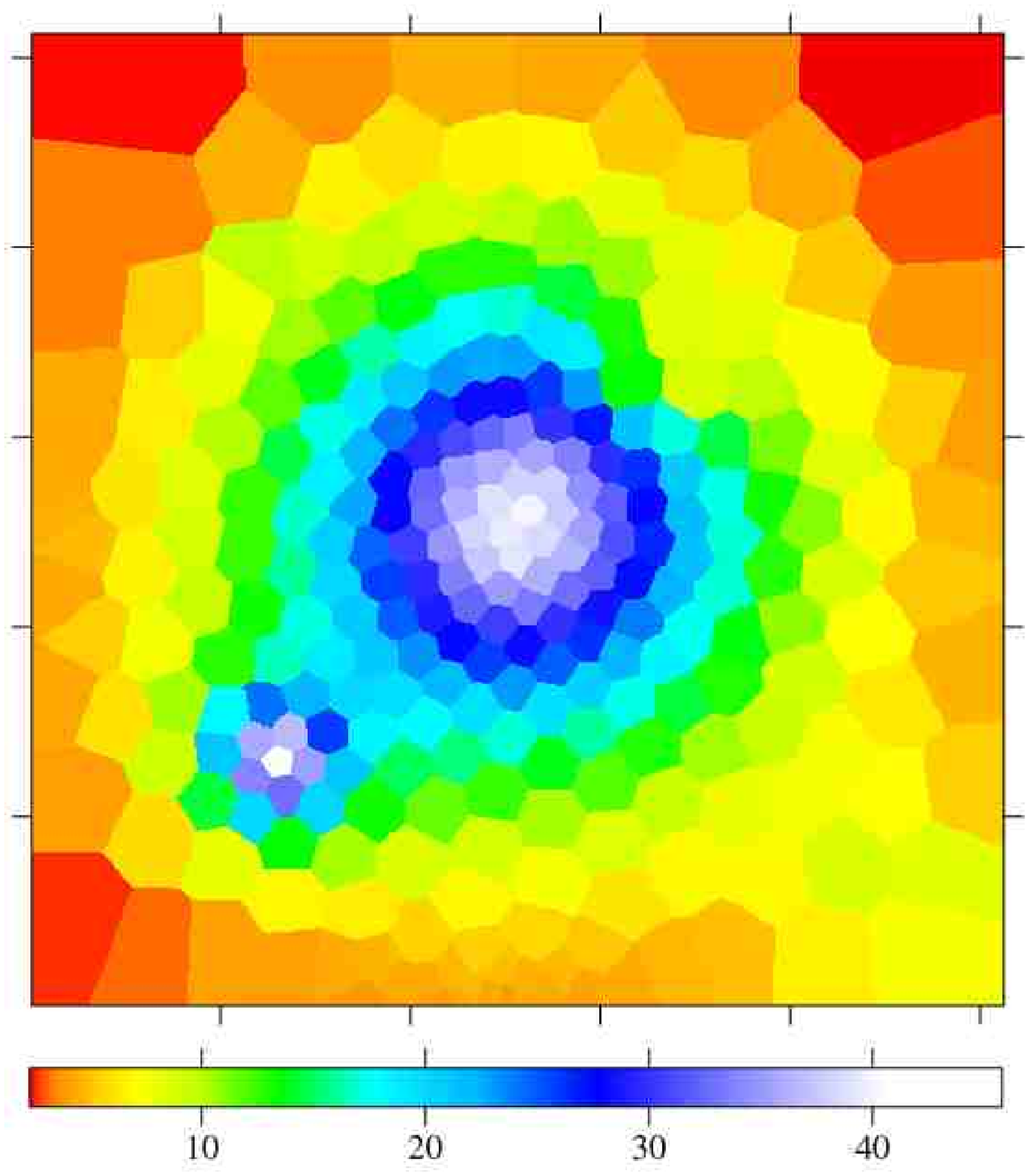}
  \includegraphics[width=0.247\textwidth]{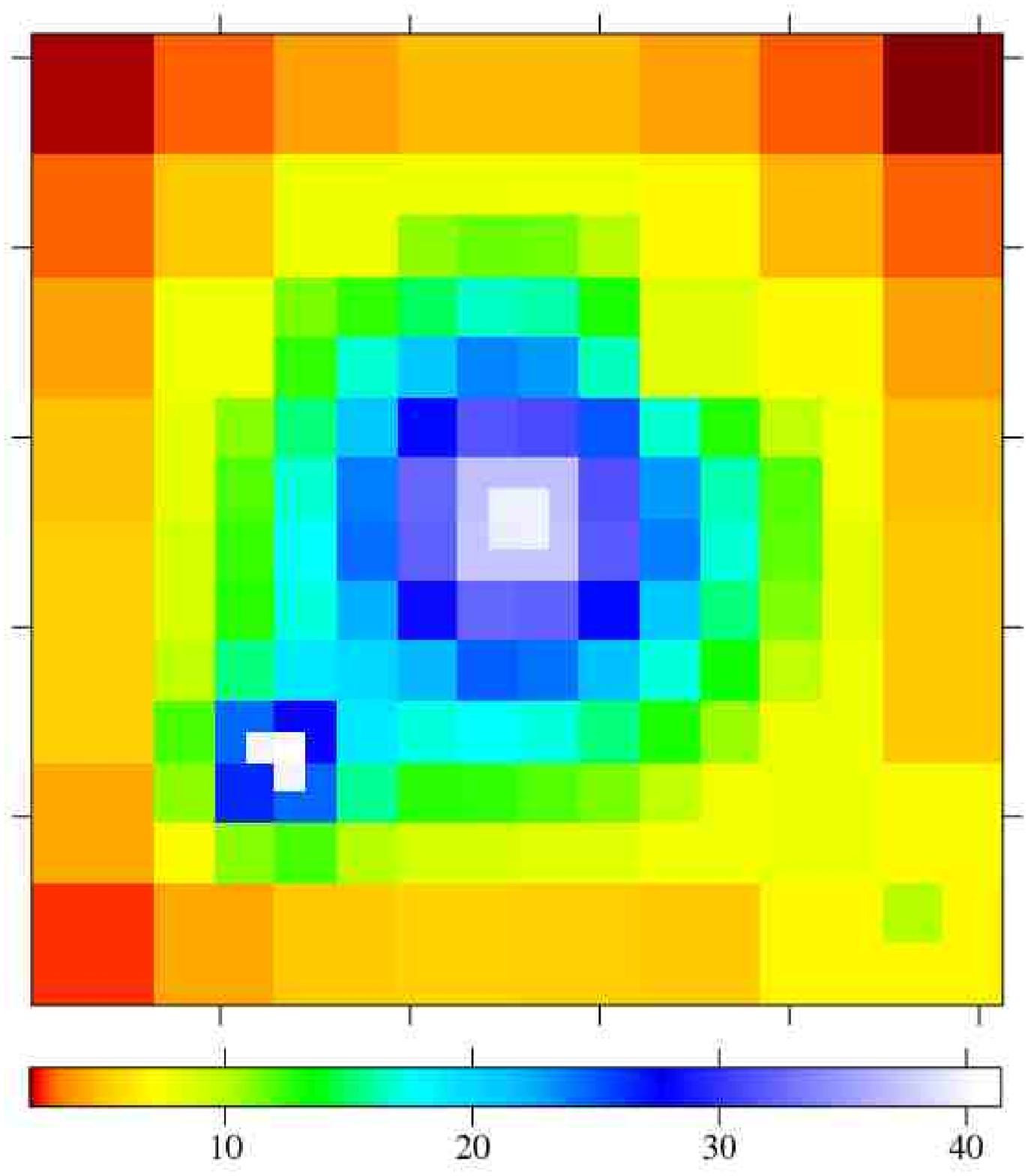}
  \caption{(Left) Contour binned surface brightness image with 
    $T=30$, $M=15$. (Centre Left) Contour binned image with $T
    = 100$, $M = 40$ and constraint $C=2$ applied. (Centre Right)
    Image created using bin-accretion algorithm of Cappellari \& Copin
    (2003) with signal-to-noise ratio of 100. (Right) Adaptively
    binned image (Sanders \& Fabian 2001) with a fractional error of
    0.01 (equivalent to a signal-to-noise ratio of 100). All images
    have units of counts per pixel and the colour scales are the same
    as in Fig.~\ref{fig:sb}.} 
  \label{fig:binned}
\end{figure*}

We binned the simulated image in Fig.~\ref{fig:sb} with the algorithm.
Fig.~\ref{fig:binned} shows the image binned using a signal-to-noise
ratio, $T$ threshold of 30 and 100 (smoothing with ratios, $M$, of
15 and 40, respectively). The $T=100$ image has also had a
constraint of $C=2$ applied. We also show a binned image using a
signal-to-noise ratio of 100 using the bin-accretion algorithm of
Cappellari \& Copin (2003), and simple adaptive binning (Sanders \&
Fabian 2002) for comparison. The contour binning method appears to
give a good reconstruction of the surface brightness. Using the same
signal-to-noise ratio, it follows the surface brightness contours
better than the tessellated image.

\begin{figure*}
  \includegraphics[width=0.247\textwidth]{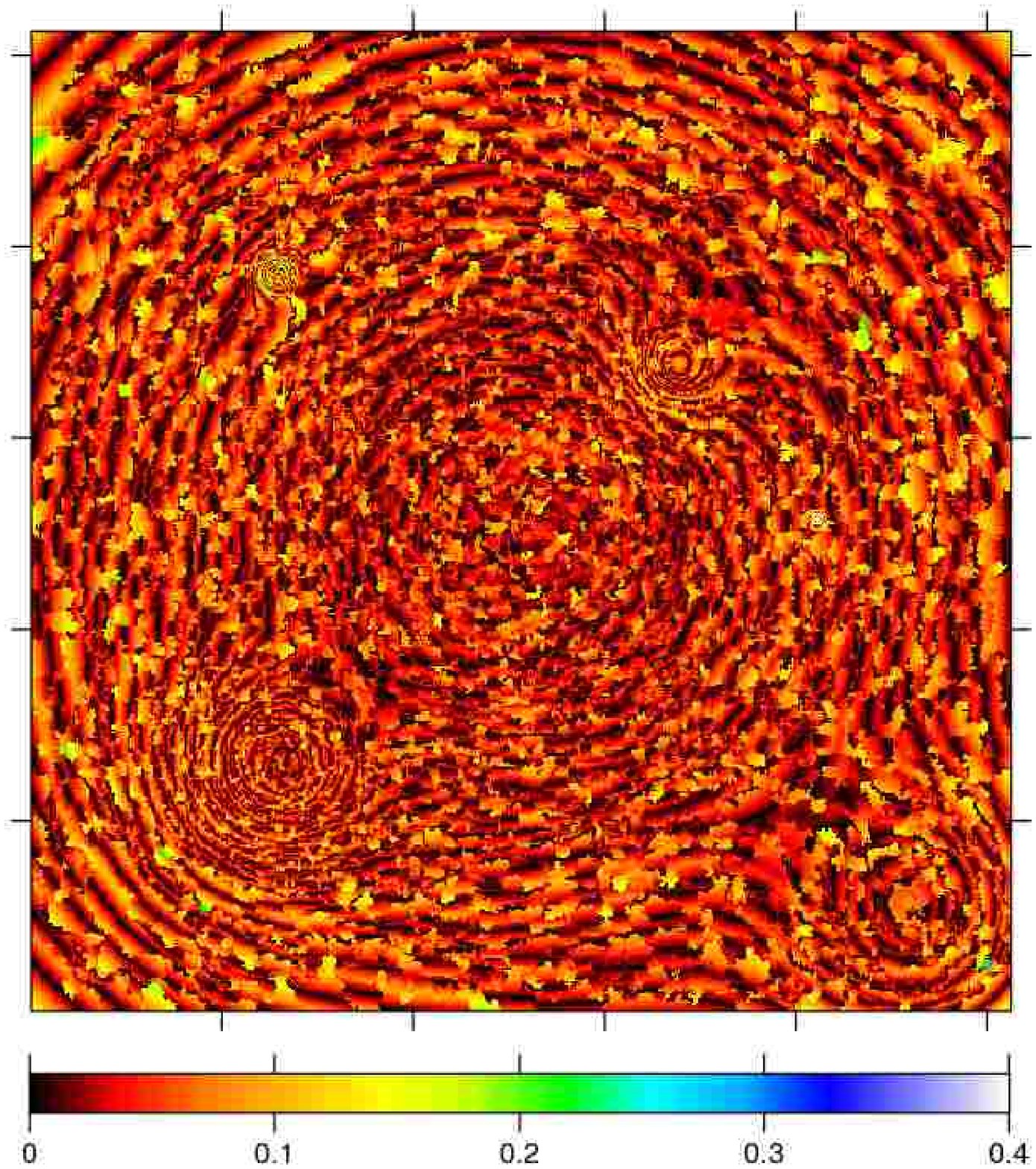}
  \includegraphics[width=0.247\textwidth]{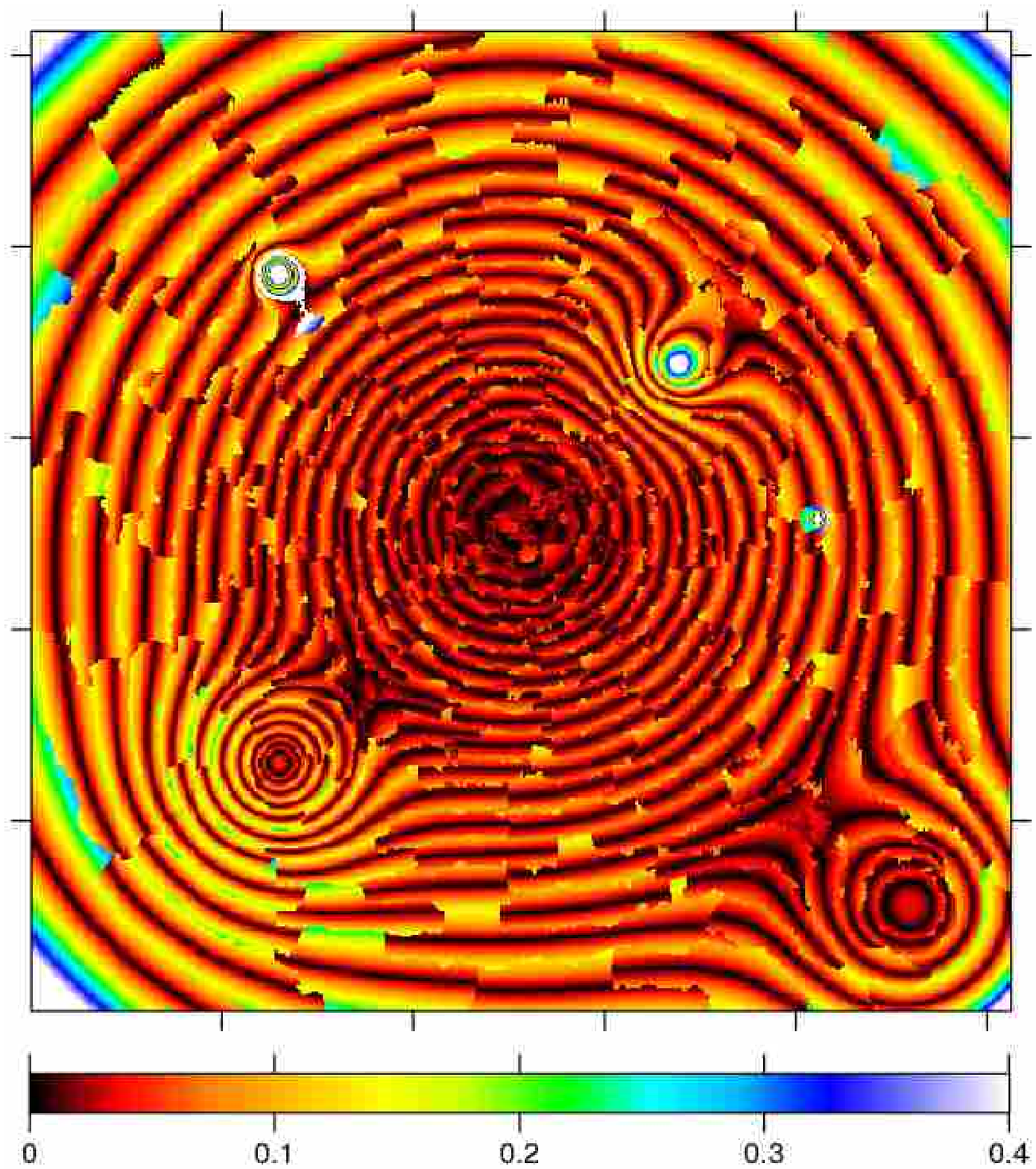}
  \includegraphics[width=0.247\textwidth]{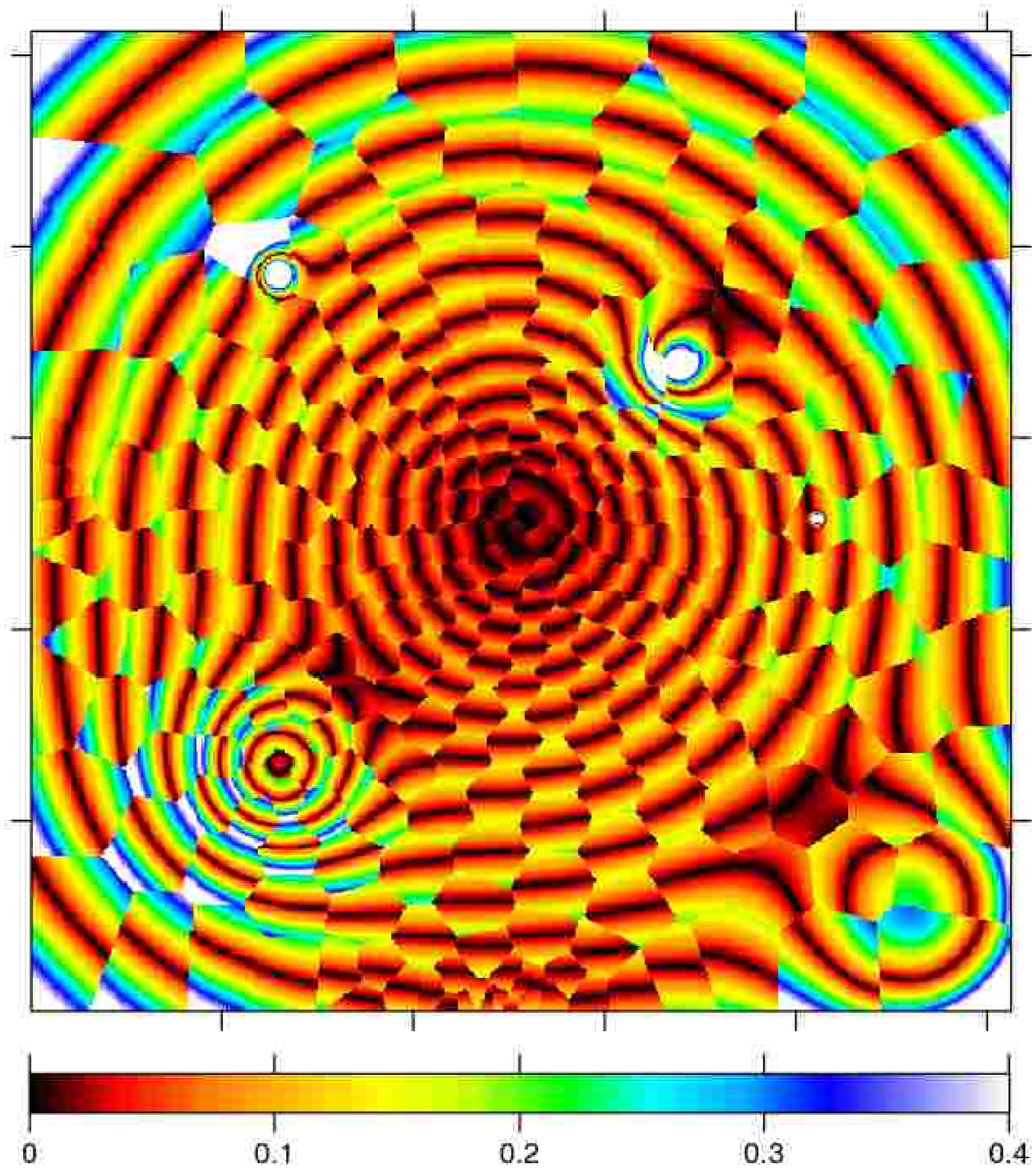}
  \includegraphics[width=0.247\textwidth]{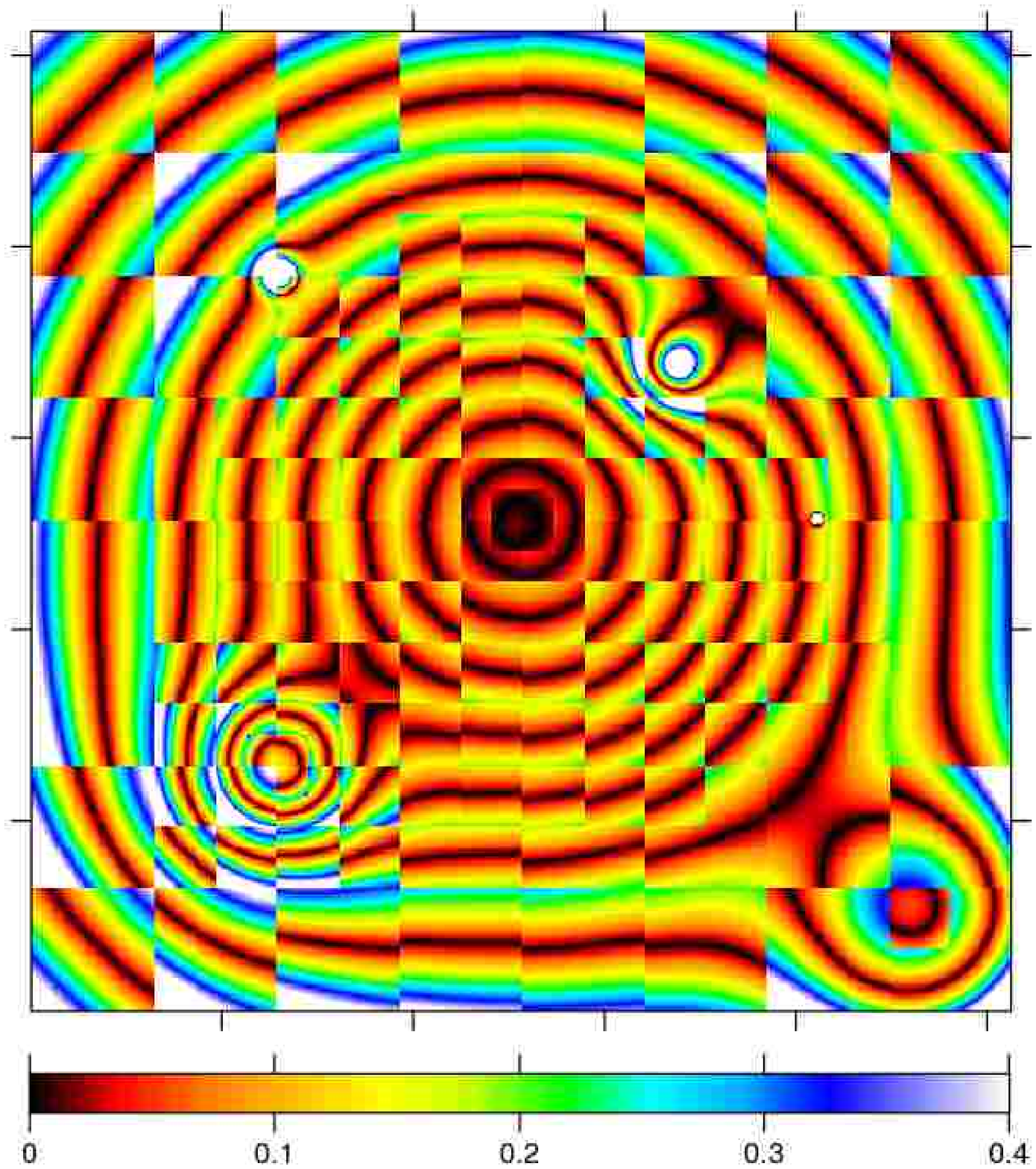}
  \caption{Absolute fractional differences between binned surface
    brightness image and model image. The panels shown are the same
    binning as Fig.~\ref{fig:binned}}
  \label{fig:binned_deltas}
\end{figure*}

In Fig.~\ref{fig:binned_deltas} we show absolute fractional difference
maps between the binned images in Fig.~\ref{fig:binned} and the model
map (Fig.~\ref{fig:sb}~left). The comparison between the contour
binning and bin accretion methods is not completely fair, as the
contour binning procedure uses a minimum signal-to-noise ratio of 100,
whilst bin accretion uses a mean signal-to-noise ratio of 100. However
increasing the signal-to-noise ratio used by the bin accretion method
would increase the size of the bins further.

\begin{figure}
  \includegraphics[width=\columnwidth]{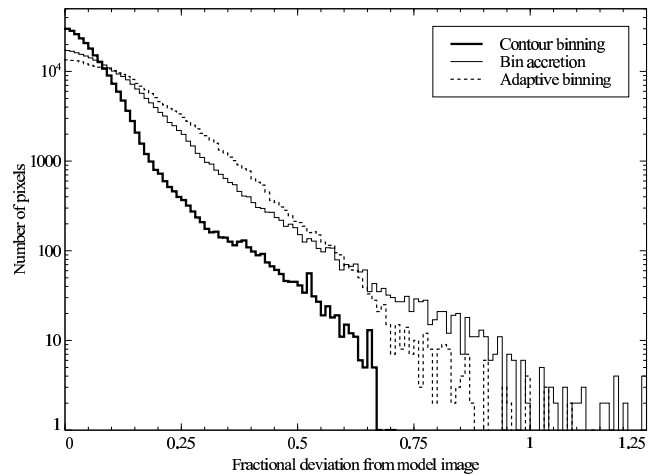}
  \caption{Plot of the distribution of absolute deviation of binned
    pixels from the model values for similar signal-to-noise ratios.
    The contour binning and adaptive binning values were taken from
    Fig.~\ref{fig:binned_deltas} (centre left and far right panels).
    The signal-to-noise ratio of the bin accretion used has been
    adjusted from 100 to 106 to get a similar number of bins compared
    to the contour binning results (257 versus 255). This makes little
    difference to the distribution. The plot shows that contour binned
    surface brightness is closer to the model surface brightness than
    the other methods.}
  \label{fig:binned_delta_dists}
\end{figure}

The signal-to-noise ratio of the bin accretion method can be adjusted
to give almost the same number of bins as contour binning. The
distribution of the deviations of pixels from the model are shown for
the three methods in Fig.~\ref{fig:binned_delta_dists}. The plot shows
that contour binning produces values which are considerably closer to
the model distribution than the other methods, due to its method of
grouping together pixels with the similar values.

\begin{figure}
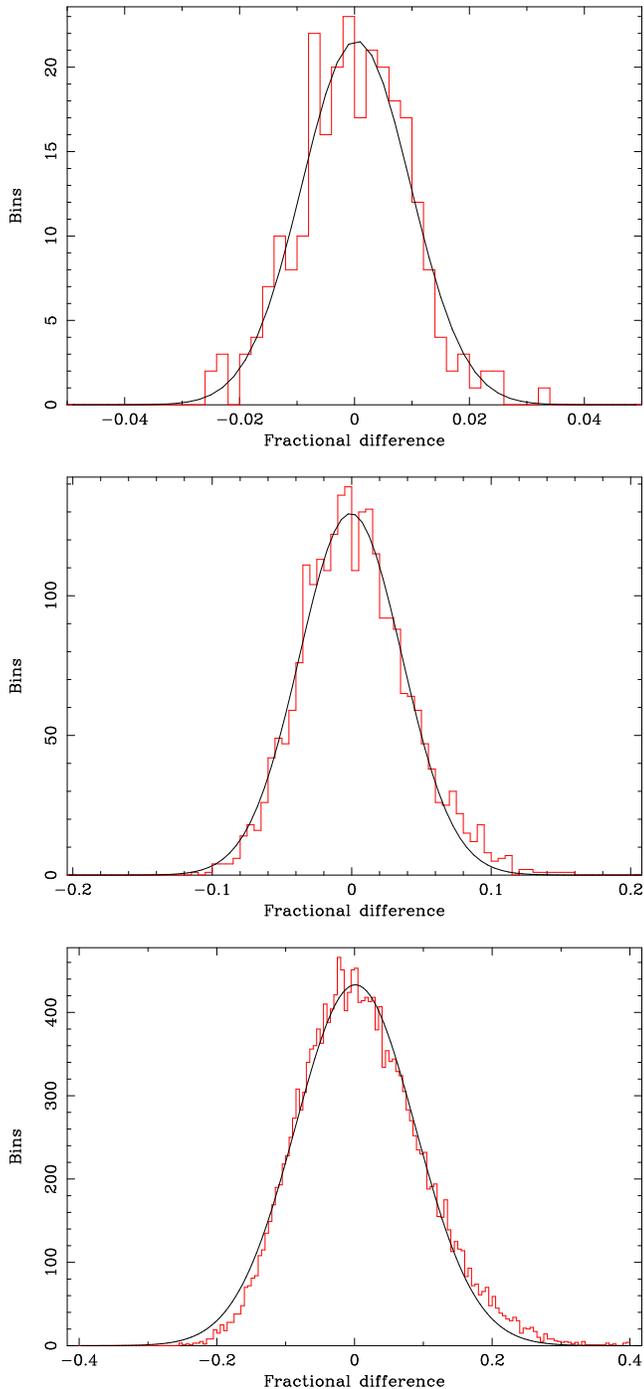

  \includegraphics[angle=270,width=\columnwidth]{fig07_1_col.eps} \vspace{0mm} \\
  \includegraphics[angle=270,width=\columnwidth]{fig07_2_col.eps} \vspace{0mm} \\
  \includegraphics[angle=270,width=\columnwidth]{fig07_3_col.eps}\\

  \caption{Distribution of fractional difference between the binned
    surface brightness of a bin and the mean model surface brightness
    in the bin.  The smooth curve is a best fitting Gaussian. (Top)
    Fractional differences for $T = 100$, $M = 40$, constraint applied
    with $C=2$. Best-fitting Gaussian centre is $3.0 \times 10^{-4}$,
    width is $9.4\times 10^{-3}$ (Centre) $T = 30$, $M = 15$. Gaussian
    parameters of ($-6.8\times 10^{-4}$, 0.037) (Bottom) $T=10$, $M =
    15$. Gaussian parameters of ($1.3\times 10^{-3}$, 0.087).}
  \label{fig:binned_delta_dist}
\end{figure}

For contour binning, the deviations appear random over the bins,
except for the two peaked sources with large signal-to-noise ratios.
In Fig.~\ref{fig:binned_delta_dist} is shown the distribution of the
fractional differences between the binned surface brightness map and
the model map for the contour binned images shown, plus another with
$T=10$. The plots show the the differences are close to Gaussian. At
the low signal-to-noise ratio of 10 there is some asymmetry in the
distribution towards higher fractional differences. This may be due to
the use of the Gehrels approximation for the Poisson errors. The width
of the distributions is close to the estimated value of $1/T$. The
estimated width of the distribution is the same as at the end of
Section~\ref{sect:acsmtest}. $c = T^2$ counts are required to exceed
the threshold without a background. The standard deviation on a mean
of $c$ counts is $\sqrt{c}$. If this is turned into a fractional width
of the distribution it equals $1/\sqrt{c} = 1/T$.

\subsection{Spectral fitting tests}
To thoroughly test the algorithm when using it to define regions to
perform spectral extraction and fitting on, we constructed simulated
\emph{Chandra} event files for a cluster of galaxies.  To produce the
files, we iterated over cells in a 3-dimensional cube of dimensions
$500 \times 500 \times 1000$ pixels, using a cell size in $(x, y, z)$
of 2, 2 and 4 pixels. In each cell we used \textsc{xspec} to fake a PI
(pulse-invariant) spectrum corresponding to the gas in the cluster
that would be in that cell. To do this we use an analytical form of
the temperature, density and abundance of the gas as a function of
radius. The density is used to calculate the emission measure, and the
spectrum is faked using the \textsc{mekal} spectral model (Mewe,
Gronenschild \& van den Oord 1985; Liedahl, Osterheld \& Goldstein
1995), with Galactic photoelectric absorption.

We added the spectra along each line of sight in $z$ to produce a
2-dimensional set of spectra. For each spectrum, we manufactured faked
events which would produce the spectrum if it were extracted from the
$2\times2$ pixel projected cell on the sky. The events were randomised
in position over the projected cell. 

In detail, a template event file from a real \emph{Chandra}
observation was populated with these events. To create the faked
spectra a constant response and ancillary-response was used in each
cell. We used a response for the centre of the ACIS-S3 CCD in 2000.
The spectra were faked for an observation of 60~ks. In addition to the
emission of the cluster in each cell on the sky, we added events which
would produce a background spectrum corresponding to the spectrum
observed from a blank-sky observation (generated by fitting a
three-component power law model to spectra from a blank-sky background
field).

To model the emission from a cluster, we fitted analytic models to the
deprojected temperature, density and abundance profiles measured in
Abell~2204 (Sanders, Fabian and Taylor 2005a). The cluster was centred
in the centre of the cube, and the properties of the gas at each
radius in the cluster was found using the analytic fits to the A2204
profiles.

The density profile in A2204 was fitted with a $\beta$ density model
of functional form
\begin{equation}
  n_e(r) = n_0 \left[ 1 + \left( \frac{r}{r_c} \right)^2
  \right]^{-3\beta /2},
\end{equation}
with $r_c = 13.97 \kpc$, $n_0 = 0.1978 \pcmcu$ and $\beta =
0.465$. The temperature was modelled with
\begin{equation}
  T(r) = T_0 + \: T_1 \frac { (r/r_c)^\eta} { 1 + (r/r_c)^\eta },
\end{equation}
with $T_0 = 2.98 \keV$, $T_1 = 7.38 \keV$, $r_c = 50.2 \kpc$ and $\eta
= 2.97$. The abundance was modelled with the same functional form,
with $Z_0 = 0.156 \Zsun$, $Z_1 = 0.921$, $r_c = 47.8 \kpc$, and $\eta
= -7.56$. The simulated cluster was assumed to lie at a redshift of
0.15, and we used a cosmology of $H_0 = 70 \kmpspMpc$ and
$\Omega_\Lambda = 0.7$. Using \emph{Chandra} pixel sizes, 1 pixel
corresponds to 1.28~kpc for the real cluster. 

Using the faked dataset, we created an image of the simulated cluster,
and of the simulated background, in the 0.5 to 7.0 keV band. This was
contour binned using a signal-to-noise of 75 ($\sim 5600$ counts),
smoothing the image with a signal-to-noise of 20, and constraining the
bins using a parameter $C=2$ (see Section~\ref{sect:constrain}). This
procedure yielded 52 bins in total.

\begin{figure*}
  \includegraphics[width=\textwidth]{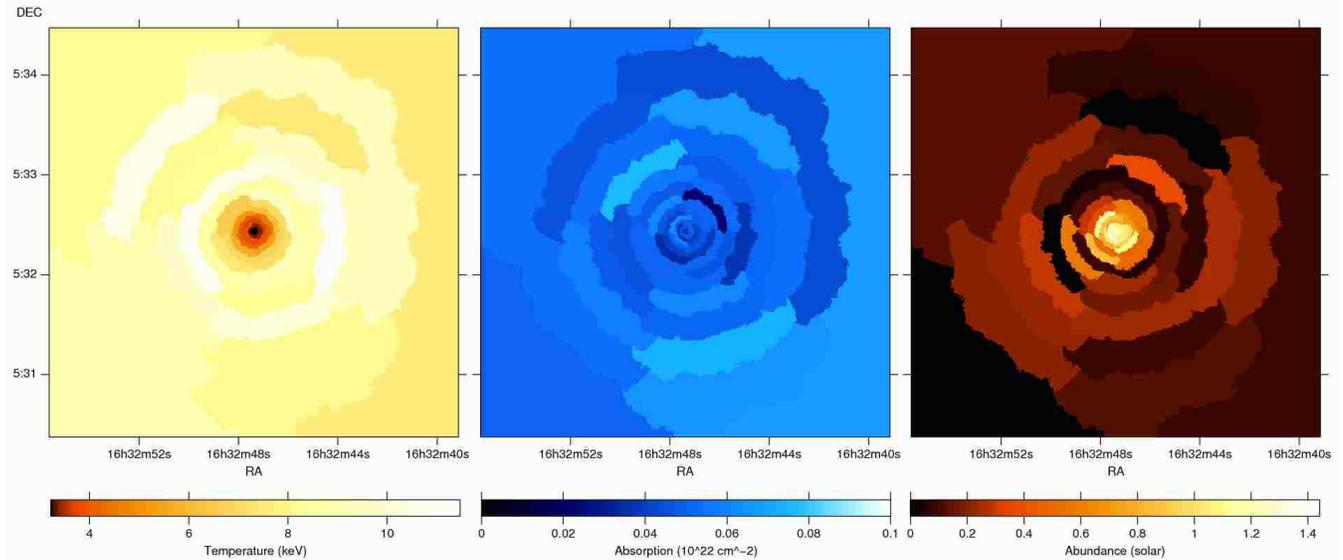}
  \caption{Resulting emission weighted temperature, photoelectric
    absorption and abundance maps generated from the simulated
    cluster. Each region was created to have a minimum signal-to-noise
    ratio of 75. Fig.~\ref{fig:simprof} shows the extracted radial
    profiles.}
  \label{fig:simmap}
\end{figure*}

\begin{figure*}
  \includegraphics[width=\textwidth]{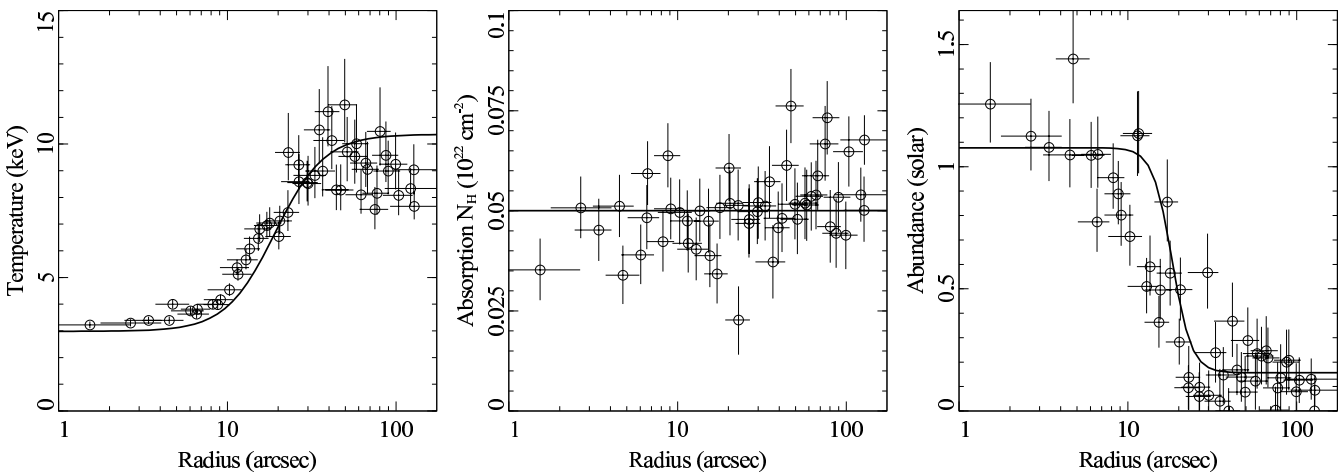}
  \caption{Profiles produced by fitting bins with a minimum
    signal-to-noise ratio of 75 using the faked data. (Left)
    temperature, (centre) photoelectric absorption and (right)
    abundance. The solid lines show the intrinsic (not projected)
    profiles the faked data was generated with.}
  \label{fig:simprof}
\end{figure*}

Individual spectra were extracted from the dataset using the regions
defined by each bin, and each spectrum was fitted with a
\textsc{mekal} model with free absorption, metallicity, temperature
and normalisation. The spectra were fit between 0.5 and 7~keV,
minimising the $\chi^2$ statistic.  We show the resulting maps in
Fig.~\ref{fig:simmap}, whilst radial profiles are shown in
Fig.~\ref{fig:simprof}.

The results show the accumulative binning algorithm does a reasonable
job at reproducing the input profiles. They do demonstrate some
deviations which are due to projection effects, and emission
weighting, however. Any method which does not account for projection
will be susceptible to these problems. In particular the cool gas in
the centre is measured to be too hot. This is due to the projection of
the hotter gas outside this region. To account for projection, you
must assume some sort of symmetry.

\begin{figure}
  \centering
  \includegraphics[width=0.7\columnwidth]{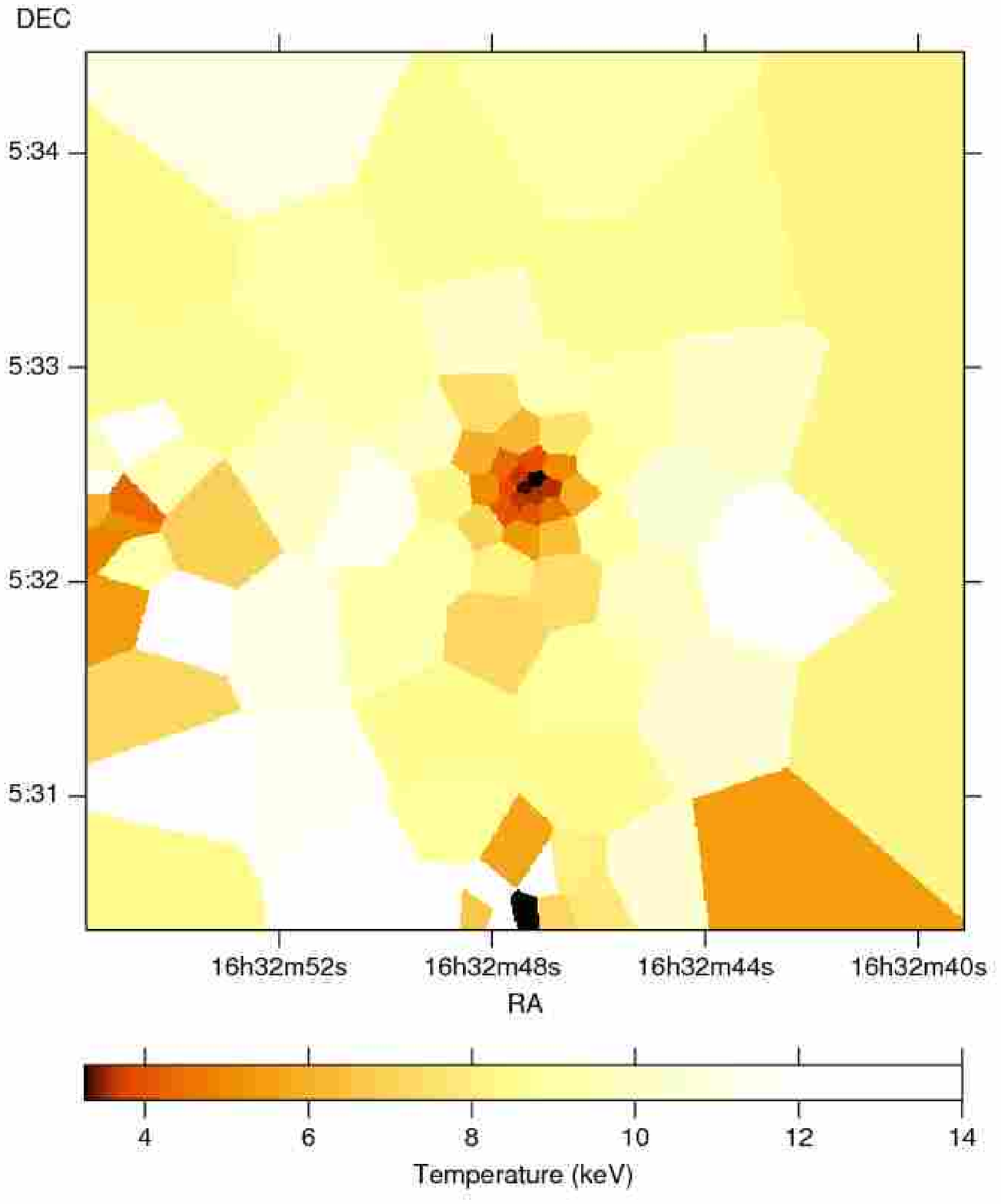}
  \caption{Temperature map created by applying bin accretion algorithm
    to fake data.}
  \label{fig:simtess}
\end{figure}

As a comparison, Fig.~\ref{fig:simtess} shows the temperature map from
the data binned using the bin accretion algorithm, with the same
signal-to-noise ratio. In the central region the bins are not as well
shaped to match the surface brightness, leading to a blocky
appearance. There are some problems with small bins being generated in
the outskirts of the image, but it is possible this could be due to
problems in our implementation of the algorithm. Again the comparison
is not quite fair as bin accretion uses a mean signal-to-nose ratio,
rather than a minimum one. However, increasing the signal-to-noise
ratio would make the bins larger.

\section{Cassiopeia A}
Cassiopeia A is a very good target to demonstrate the algorithm on.
The large range in brightness of the image between the bright knots
and the darker regions shows off the advantages of this particular
method. We present here results using a 49.7~ks observation of Cas~A
using the ACIS-S3 detector on \emph{Chandra} taken on 2002-02-06
(observation number 1952).

\subsection{Data preparation}
Firstly we reprocessed the level 1 event file using \textsc{ciao}
3.3.0.1 with the gain file acisD2000-08-12gainN0005.fits. Time
dependent gain correction was also applied using the
\textsc{corr\_tgain} utility using the November 2003 correction
(Vikhlinin 2003). In addition the \textsc{ciao} version of the
\textsc{lc\_clean} script (Markevitch 2004) was used to remove periods
of the dataset where the count rate was 20 per~cent away from the
mean. Very little time was removed, producing a level 2 event file
with an exposure of 48.3~ks, although such filtering is difficult on
such a bright target.

The dataset was taken using the unusual GRADED observation mode of the
ACIS detector, where events likely to be background are discarded at
the satellite rather than removed by the observer. Despite this, the
background rate at high energy (9~keV or more) matches that from blank
sky observations. We used a tailored version of a blank sky
observation to account for background in this analysis.

As Cas~A is a very bright object it is important to take account of
`out of time' events. These are events which occur whilst the detector
is being read out. As the ACIS detectors use framebuffers this time is
short. Around 1.3~per~cent of events occur out of time. We used the
\textsc{make\_readout\_bg} script (Markevitch 2003) to generate a
synthetic event file to include in the background subtraction. The
script works by randomising the CHIPY coordinate of the events in the
input event file (which is the actual observation), and increasing the
exposure time by the reciprocal of out of time fraction. In this
observation the out of time background dominates over the blank sky
background over much of the energy range. 

\subsection{X-ray image}
\begin{figure}
  \includegraphics[width=\columnwidth]{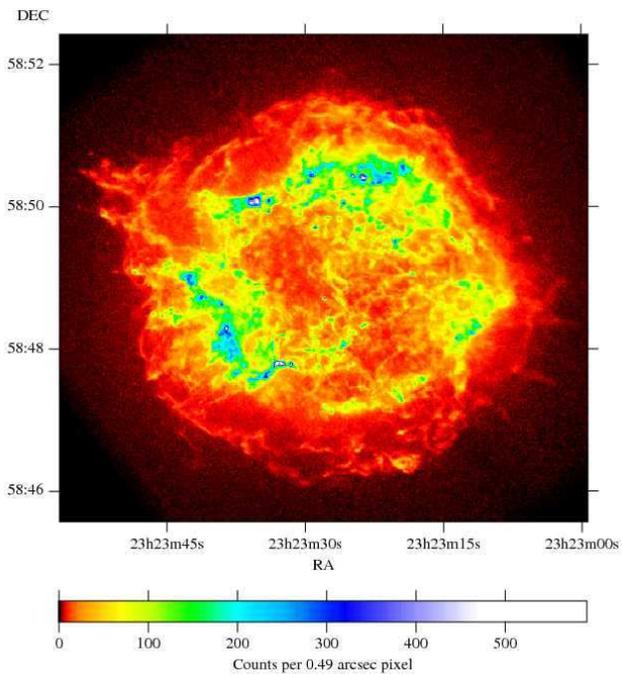}
  \caption{Background subtracted, accumulatively smoothed image of
    Cas~A in the 0.5 to 7~keV band, smoothed using a signal-to-noise
    ratio threshold, $M$ of 15.}
  \label{fig:cassa_img}
\end{figure}

\begin{figure*}
  \includegraphics[width=\textwidth]{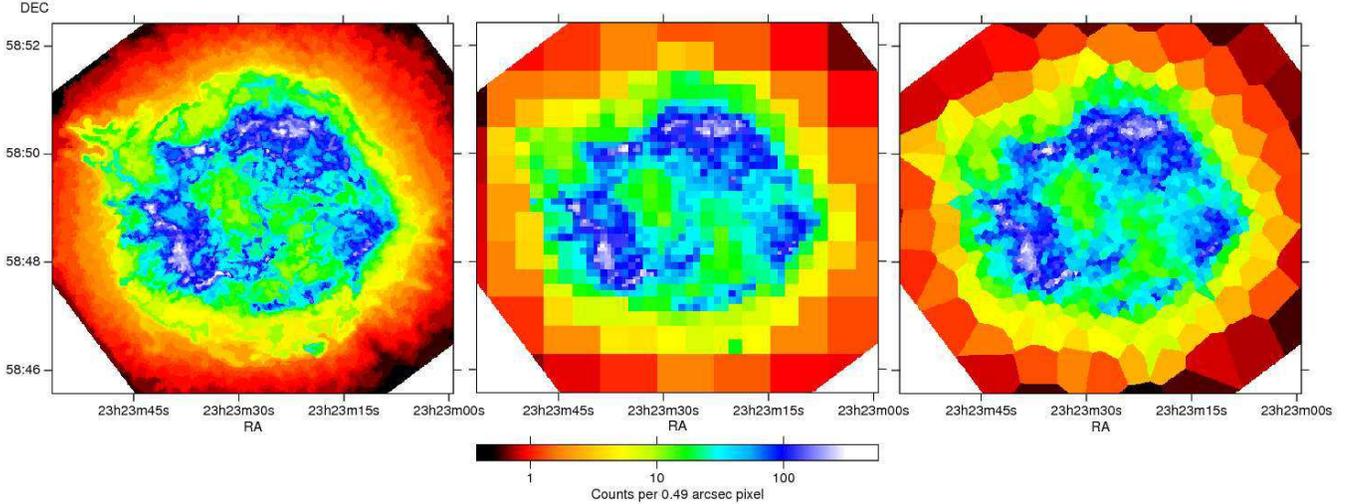}
  \caption{Binned images of Cas~A. (Left) Contour binned image using
    a signal-to-noise binning threshold ratio, $T$, of 100. The image
    was smoothed using $M=15$ before binning. (Centre) Adaptively
    binned image with a fractional error threshold of 0.01 using
    algorithm of Sanders \& Fabian (2001).  Note that this binning did
    not take account of the readout background.  (Right) Bin accretion
    binned image with a signal-to-noise of 100 using algorithm of
    Cappellari \& Copin (2003).}
  \label{fig:cassa_binned}
\end{figure*}

\begin{figure}
  \includegraphics[width=\columnwidth]{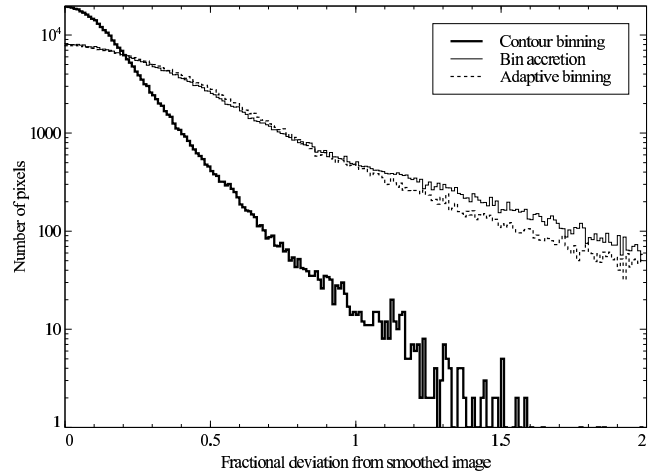}
  \caption{Histogram of fractional deviations of each pixel of a
    binned image from accumulatively smoothed image of Cas~A (signal
    to noise threshold of 8). The area examined was limited to a 2.8
    arcmin radius around the core of the remnant. Shown are the
    results for contour binning, adaptive binning and bin accretion
    using similar signal-to-noise ratios.}
  \label{fig:cassa_deltas}
\end{figure}

In Fig.~\ref{fig:cassa_img} is shown an accumulatively smoothed X-ray
image of the dataset. We show in Fig.~\ref{fig:cassa_binned} a contour
binned image of the dataset with a signal-to-noise threshold, $T$, of
100 (the image was smoothed with a signal-to-noise threshold of 15
before binning) along with adaptively and accretion binned images with
similar signal-to-noise ratios.  The binning procedure created 1176
spatial bins, giving an average of $1.2 \times 10^4$ counts per
spectrum before background subtraction.  It can be seen that the bins
follow the shape of the surface brightness well, and better than the
other binning methods. Note that there is relatively bright emission
around the remnant (this is difficult to see in plots shown, due to
the greyscale colour scaling). This emission is in fact radiation from
Cas~A scattered by dust in the interstellar medium. The measurements
presented later include results outside of the remnant which are from
the scattered radiation.

To compare the binning methods better, we can compare the surface
brightness of binned pixels against what they are expected to be.  If
we take the absolute fractional difference between a binned
accumulately smoothed image of the remnant and an accumulatively
smoothed image, then plot the histogram of the differences, we obtain
the distributions in Fig.~\ref{fig:cassa_deltas}. The area examined
was limited to a radius of 2.8~arcmin from the centre of the remnant
to avoid edge effects.  The plot shows that contour binning is much
better at reproducing the accumulately smoothed surface brightness of
the object than the other methods. This argument assumes that an
accumulative smoothed image is a good estimate of the intrinsic sky
surface brightness. To make the comparison fairer we increased the
signal-to-noise used by the bin accretion method to 109 to match the
number of bins used by contour binning, although it makes very little
difference to the results. The number of bins used by contour binning,
bin accretion and adaptive binning are 1175, 1182 and 889
respectively. Contour binning also produces the most accurate surface
brightnesses from our simulated data
(Fig.~\ref{fig:binned_delta_dists}). The test with the simulated data
is less biased as we know the intrinsic surface brightness of the
object, although the results are similar.

\subsection{Spectral fitting}
We extracted spectra from the events file in each of the binned
regions. Rather than use the standard \textsc{ciao} \textsc{dmextract}
tool we used our own code which extracts all the spectra in a single
pass. The spectra were grouped to have a minimum number of 20 counts
in each spectral channel using \textsc{grppha} from \textsc{ftools}. 
Using the \textsc{ciao} \textsc{mkwarf} and \textsc{mkrmf} tools, we
generated weighted response and ancillary response files for each of
the regions, weighted by the number of counts between 0.5 and 7 keV. 
In addition we extracted background spectra from the blank sky and out
of time event files. 

The spectra were fit using \textsc{xspec} 11.3.2 between energies of
0.6 and 7 keV. We fitted a \textsc{vnei} model to each region, with
the neivers parameter set to 2.0, which uses the ionisation fractions
of Mazzotta et al (1988) and uses \textsc{aped} (Smith et al 2001) to
generate the resulting spectrum. The emission model was absorbed by a
\textsc{phabs} absorption model (Balucinska-Church \& McCammon 1992).
In the fit the absorption, temperature, normalisation, ionisation
timescale, redshift, Ne, Mg, Si, S, Ar, Ca, Fe and Ni abundances were
free parameters (using a maximum of $40 \Zsun$ in each abundance
parameter). The H, He, C, N and O abundances were fixed at solar
values.  We used the solar abundance ratios of Anders \& Grevesse
(1989).  In some regions there are prominent lines at Fe-K energies
(e.g. Hwang et al 2000), which lead to poor reduced $\chi^2$ values.
We therefore added a Gaussian component to the model allowed to vary
between 6 and 7 keV.

With the large number of free parameters it was easy for the fits to
become stuck in local minima. We tried to avoid this by freeing the
most influential parameters first, and doing a fit after each freeing. 
Also during the fit we searched for errors on the parameters to help
find other $\chi^2$ minima. However, this prescription does not
guarantee to find the absolute minimum in $\chi^2$ space, so some
parameters may not be at their best-fitting values. In addition the
best fitting parameters may not be physically sensible. After the
spectra were fit, the best-fitting values of each parameter were used
to generate maps.

The model used is somewhat simplistic compared to those used by other
authors, not taking account of multi-temperature regions or allowing
the lines from each element to have different velocities (except for
Fe-K). Most of the features reported by others are reproduced,
however, as we discuss below. The model also assumes the emission is
thermal in origin, and does not include a non-thermal component.
Hughes et al (2000) have noted continuum dominated regions. These
regions can be fitted by a powerlaw component or an absorbed
bremsstrahlung.  Hughes et al (2000) prefer the interpretation that
the continuum is X-ray synchrotron emission.

\begin{figure*}
  \includegraphics[width=0.7\textwidth]{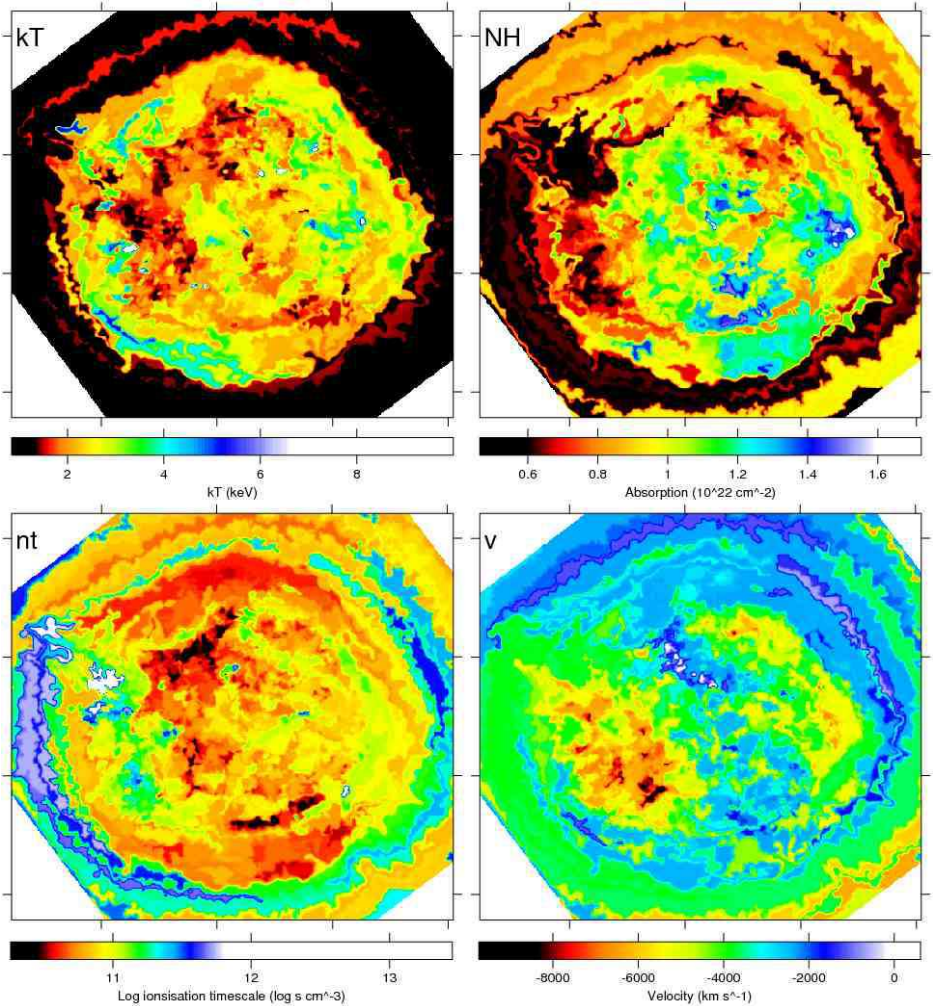}
  \caption{Temperature, absorbing column density, ionisation timescale
    and fitted velocity in Cas~A.}
  \label{fig:cassmisc1}
\end{figure*}

\begin{figure*}
  \includegraphics[width=0.7\textwidth]{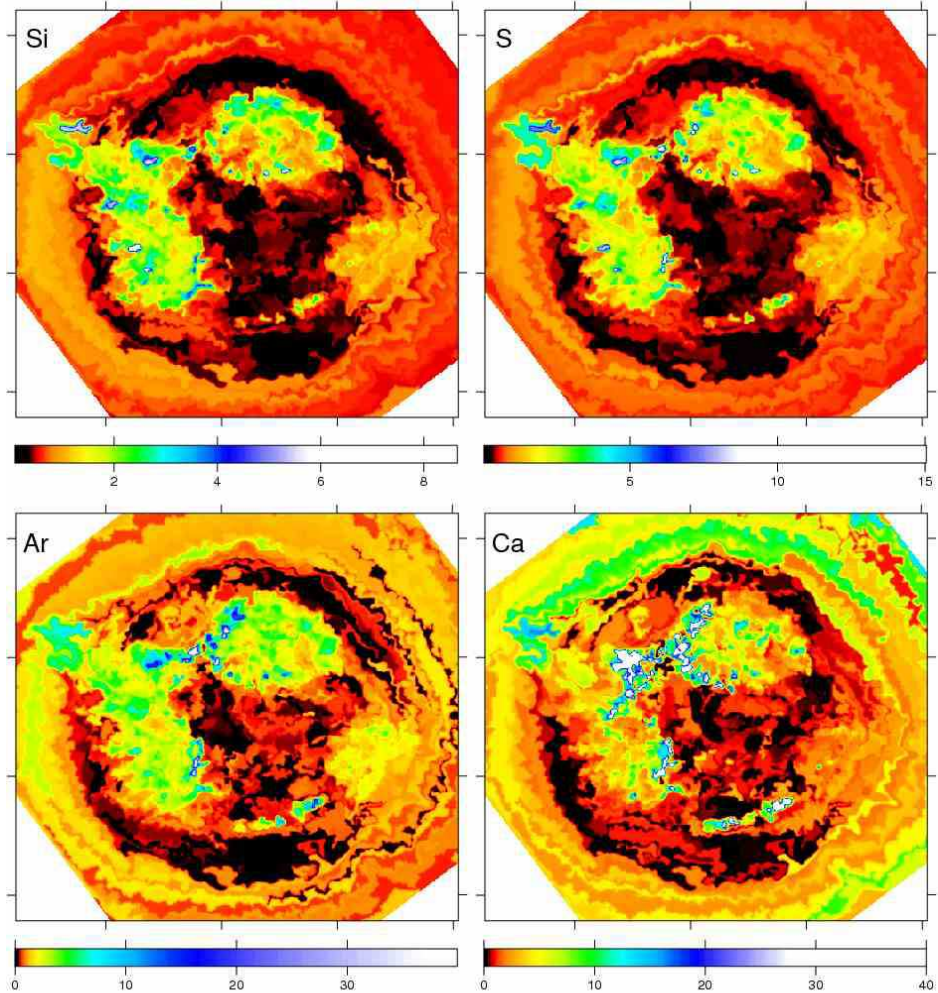}
  \caption{Abundances in Cas~A of Si, S, Ar and Ca relative to solar.}
  \label{fig:cassSiZ}
\end{figure*}

\begin{figure*}
  \includegraphics[width=0.7\textwidth]{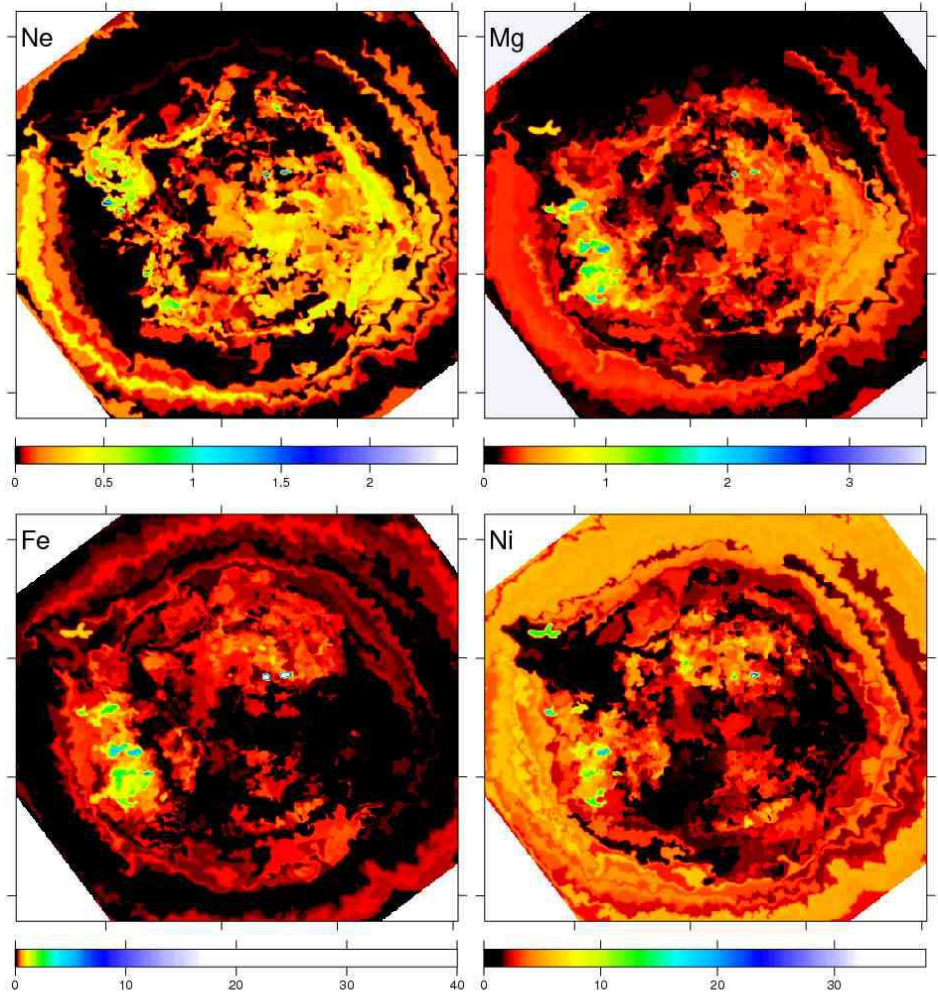}
  \caption{Abundances in Cas~A of Ne, Mg, Fe and Ni relative to
    solar.}
  \label{fig:cassFeZ}
\end{figure*}

\begin{figure*}
  \includegraphics[width=0.7\textwidth]{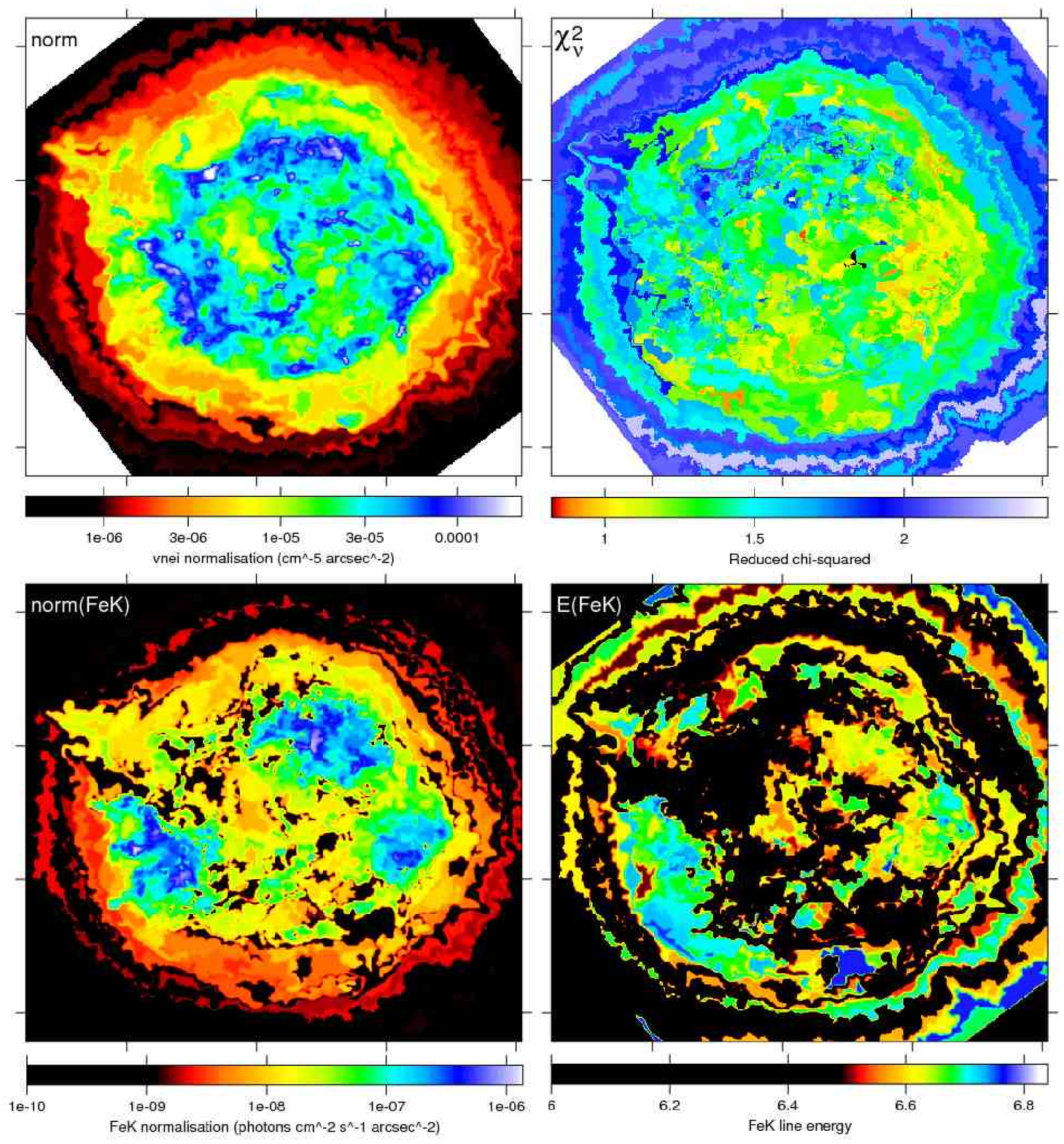}
  \caption{\textsc{vnei} normalisation per unit area (a proxy for the
    emission when the line emission has been removed), reduced
    $\chi^2$ of the fit, normalisation of the Fe-K line per unit area,
    and energy of the Fe-K line in Cas~A.}
  \label{fig:cassmisc2}
\end{figure*}

In Fig.~\ref{fig:cassmisc1} are shown the best fitting temperature,
absorbing column density, ionisation timescale and velocity. The
velocity was obtained by multiplying the best fitting redshift by the
speed of light. Abundances for Si-like elements are shown in
Fig.~\ref{fig:cassSiZ}, and for Fe-like elements in
Fig.~\ref{fig:cassFeZ}. We finally display in Fig.~\ref{fig:cassmisc2}
the normalisation of the main \textsc{vnei} component per unit area,
the reduced-$\chi^2$ of the fits, the normalisation of the Gaussian
Fe-K component and its energy.  All the maps here have been smoothed
with a Gaussian of 2 pixels (0.98 arcsec) for display purposes. The
positions in the map and text are in J2000 coordinates.

\subsection{Discussion}
The absorption map (Fig.~\ref{fig:cassmisc1} [top right]) shows a
dense clump of X-ray absorbing material to the west of the nebula in
great detail. This material was itself found by previous observations
(Keohane, Rudnick \& Anderson 1996; Willingale et al 2002).

The temperature map shows the location of the inner and outer shocks
well (Gotthelf et al 2001), as regions of high temperature, in
addition to the direction of the north-east jet. The spectral fits
also indication the position of the south-west jet (Hwang et al 2004)
as higher temperature regions.

The abundance maps can be compared against those created by plotting
the ratio of emission in a band crafted to the emission line of a
particular element to continuum emission (Hwang, Holt \& Petre 2000).
This technique can only be used for elements which produce lines
easily identifiable from the spectrum, however. Our maps show good
agreement with the morphology of the ratio maps, although are some
point-to-point differences. For example, we find more emission from S
and Ar where the northern enhancement in abundance meets the eastern
region, around ($23^\mathrm{h}33^\mathrm{m}33^\mathrm{s}$ $+48^\circ
50' 05''$). In addition, although the Hwang et al (2000) maps show a
rim of Si (and possible S) enhancement to the south of the object, we
see the rim break up into separate peaks (as found in the megasecond
observation of Hwang et al 2004), and see the enhancement in S, Ar,
Ca, less strongly in Fe, and possibly in Mg and Ne. Again Si, S, Ar
and Ca vary much more uniformly than Ne, Mg, Fe and Ni.

Although we do not allow separate velocities for the different metal
components (except Fe-K), the velocity map (Fig.~\ref{fig:cassmisc1}
bottom right), where velocities towards the observer are shown as
negative, shows similar features to those found by others for the Si
line which is usually dominant (e.g.  Willingale et al 2002). We are
able to map the velocity in those regions where the line emission is
weak, as the emission is binned over larger regions. The absolute
velocity normalisation appears to be uncertain, as earlier calibration
versions with the same data gave velocities systematically around
$3500\kmps$ larger. Using the Fe-K line energy
(Fig.~\ref{fig:cassmisc2} bottom right), the trend of velocity is
again very similar to the Willingale et al.  results, but here there
is not enough signal to map the energy in low intensity regions.

The plot of the normalisation of the \textsc{vnei} component is
particularly striking (Fig.~\ref{fig:cassmisc2} top left). It
shows the emission from the remnant when the line emission has been
removed. Cas~A is much more symmetrical in the underlying emission.

The reduced-$\chi^2$ is not too bad generally when the Fe-K emission
has been modelled (as it has here), given the complexity of the
spectra and the simplicity of the fitted model. Where the
reduced-$\chi^2$ is at its worst, the model seems unable to take
account of the continuum and the emission lines together. Adding a
second temperature component in these regions improves the quality of
the fit substantially. Multiple temperature fits have been necessary
before in the spectral fitting (e.g. Willingale et al 2002). There is
much room for further investigation of the spectra here, but it is not
the aim of this paper to study the physics of these regions in detail.

\section{Conclusions}
We have shown that the accumulative smoothing method, or that used by
\textsc{fadapt}, is a robust way to estimate the real surface
brightness distribution of an extended object. We have discussed the
method in the cases of varying exposures, and including backgrounds.

We have described a new method for choosing regions for spectral
extraction, or colour analysis, called contour binning. We have shown
that method reliably creates bins which follow the surface
brightness. We demonstrate the method using a simulated cluster
dataset, and apply it to a \emph{Chandra} observation of the supernova
remnant Cassiopeia A.

The algorithm is ideal where spectral changes are associated with
changes in surface brightness, as is often the case in X-ray
astronomy. The results on objects with fine spatial detail are very
closely matched to the object, aesthetically pleasing, and easy to
interpret, compared to other methods (as exemplified by
Fig.~\ref{fig:cassa_binned} and Fig.~\ref{fig:cassa_deltas}). As the
algorithm follows the surface brightness it naturally bins regions
which are likely to be physically associated. It avoids mixing spectra
together for physically disassociated regions.

\section*{Availability}
A C++ implementation of the algorithm can be downloaded from
\url{http://www-xray.ast.cam.ac.uk/papers/contbin/} , including other
helpful associated programs.

\section*{Acknowledgements}
The author is grateful to A.C. Fabian and R.M. Johnstone for
discussions.

\clearpage

\end{document}